\documentclass[final,5p,times,twocolumn,authoryear]{elsarticle}
\usepackage[colorlinks=true,citecolor=blue]{hyperref}
\usepackage{rotating}

\usepackage{amssymb}
\usepackage{lipsum}
\usepackage{adjustbox}
\usepackage{longtable,lscape}
\usepackage{booktabs}
\usepackage{url}
\usepackage{newtxtext,newtxmath}
\usepackage{graphicx, graphics,color}
\usepackage{xcolor}
\usepackage{soul}
\usepackage[T1]{fontenc}
\usepackage[flushleft]{threeparttable}

\newcommand\kms{\ifmmode {\rm~km\ s}^{-1} \else ~km s$^{-1}$\fi}
\newcommand\Hunit{\ifmmode {\rm~km\ s}^{-1}\ {\rm Mpc}^{-1}
        \else ~km s$^{-1}$ Mpc$^{-1}$\fi}
\newcommand\ctssec{\ifmmode {\rm~count\ s}^{-1} \else ~count s$^{-1}$\fi}
\newcommand\ergsec{\ifmmode {\rm~erg\ s}^{-1} \else
        ~erg s$^{-1}$\fi}
\newcommand\funit{\ifmmode {\rm~erg\ s}^{-1}\;{\rm cm}^{-2} \else
        ~ergs s$^{-1}$ cm$^{-2}$\fi}
\newcommand\phflux{\ifmmode {\rm~photon\ s}^{-1}\;{\rm cm}^{-2}
        \else   ~photon s$^{-1}$ cm$^{-2}$\fi}
\newcommand\efluxA{\ifmmode {\rm~erg\ s}^{-1}\;{\rm cm}^{-2}\;{\rm
        \AA}^{-1} \else ~erg s$^{-1}$ cm$^{-2}$ \AA$^{-1}$\fi}
\newcommand\efluxHz{\ifmmode {\rm~erg\ s}^{-1}\;{\rm cm}^{-2}\;{\rm
        Hz}^{-1} \else ~erg s$^{-1}$ cm$^{-2}$ Hz$^{-1}$\fi}
\newcommand\cc{\ifmmode {\rm~cm}^{-3} \else cm$^{-3}$\fi}
\newcommand\FWHM{\ifmmode {\rm~FWHM} \else ${\rm~FWHM}$\fi}
\newcommand\Zsun{\ifmmode Z_{\odot} \else $Z_{\odot}$\fi}
\newcommand\Lsun{\ifmmode L_{\odot} \else $L_{\odot}$\fi}

\newcommand\hbeta{\ifmmode {\rm H}\beta \else H$\beta$\fi}
\newcommand\Kalpha{\ifmmode {\rm K}\alpha \else K$\alpha$\fi}
\newcommand\nh{\ifmmode N_{\rm H} \else N$_{\rm H}$\fi}

\usepackage{amsmath}	% Advanced maths commands
\usepackage{caption}
\usepackage{subcaption}
\usepackage[normalem]{ulem}
\usepackage{scrextend}
\usepackage[utf8]{inputenc}
\usepackage[T1]{fontenc}

\journal{JHEAP}

\begin{document}

\begin{frontmatter}



\title{Comprehensive X-ray Study of Giant Radio Galaxies with \textit{XMM-Newton} and \textit{Chandra}}

\author[first]{Niraj Maurya}
\affiliation[first]{organization={Dayanand Science College},
            addressline={Barshi Road}, 
            city={Latur},
            postcode={413512}, 
            state={Maharashtra},
            country={India}}

\author[second]{Satish Sonkamble\corref{cor1}}
\ead{satish04apr@gmail.com}

\affiliation[second]{organization={Centre for Space Research, North-West University},
            city={Potchefstroom},
            postcode={2520}, 
            state={North West Province},
            country={South Africa}}

\author[third]{Suraj Dhiwar}
\affiliation[third]{organization={Aryabhatta Research Institute of Observational Sciences (ARIES)},
            city={Nainital},
            postcode={263001},
            state={Uttarakhand},
            country={India}}

\author[first]{M.~B.~Pandge\corref{cor1}}
\ead{mbpandge@gmail.com}

\author[second,fourth]{S.~Ilani~Loubser}

\author[first]{Avinash Kale}

\affiliation[fourth]{organization={National Institute for Theoretical and Computational Sciences (NITheCS)},
            city={Potchefstroom},
            postcode={2520}, 
            state={North West Province},
            country={South Africa}}

\cortext[cor1]{Corresponding authors : }

\begin{abstract}
We present a systematic X-ray spectral analysis of 27 giant radio galaxies (GRGs) using archival \textit{XMM-Newton} and \textit{Chandra} data. Roughly 44\% of the sample show intrinsic absorption ($N_{\mathrm{H,int}} \gtrsim 10^{21}\,\mathrm{cm}^{-2}$), while several exhibit narrow Fe K$\alpha$ emission lines with equivalent widths up to $\sim$570 eV. Soft X-ray features are also common, including thermal plasma emission (kT$\sim 0.2$--0.8 keV) and ionized absorbers, in some cases consistent with high-velocity outflows. The photon indices are typically hard (median $ \Gamma \sim 1.6$), in line with radio-loud AGN. A comparison between nuclear X-ray luminosity and extended radio power shows that many sources host relatively strong X-ray cores compared to their lobes, pointing toward possible restarted nuclear activity. For 10 GRGs with published black hole masses, we explore preliminary relations with X-ray luminosity, Eddington ratio, and photon index. We find a positive trend between $M_{\rm BH}$ and $L_{2-10\,{\rm keV}}$, while Eddington ratios show no clear dependence on mass. A tentative negative relation between $\lambda_{\rm Edd}$ and $\Gamma$ hints at inefficient accretion, although confirmation will require larger samples. Overall, GRGs display diverse nuclear conditions and accretion states, supporting scenarios where obscuration, episodic or restarted activity, and low-efficiency accretion shape their long-term evolution.

\end{abstract}




\end{frontmatter}

%\tableofcontents



%%%%%%%%%%%%%%%%% BODY OF PAPER %%%%%%%%%%%%%%%%%%

\section{Introduction}
\label{intro}

Giant radio galaxies (GRGs) are radio-loud AGN whose projected linear sizes exceed 0.7 Mpc, placing them among the largest single astrophysical entity known \citep{1974Natur.250..625W,refId0,2020A&A...642A.153D,2023A&A...672A.163O,2025A&A...696A..97D,2026A&A...706A.310M}. These structures can extend up to $\sim$ 5 Mpc in physical size \citep{1998A&A...329..431M, 2008ApJ...679..149M, 2015MNRAS.449..955M, 2024Natur.633..537O}. Radio-loud galaxies have historically been classified into two morphological categories: Fanaroff–Riley Type I (FR I) and Fanaroff–Riley Type II (FR II), according to the distribution of their radio brightness \citep{1974MNRAS.167P..31F}. FR I sources are typically less luminous and show their brightest radio emission near the core, with jets that gradually fade and decelerate as they extend outward. In contrast, the more powerful FR II galaxies exhibit edge-brightened lobes, where relativistic jets propagate over large distances and terminate in prominent hotspots far from the nucleus. Although millions of radio galaxies have been identified in the past few decades, only a small fraction, on the order of a few hundred, qualify as GRGs \citep{2020yCat..36350005D}.

GRGs are believed to originate from the AGN at the core of their host galaxies, where supermassive black holes (SMBHs) with masses in the range 10$^8$ – 10$^{10}$ M$_\odot$ \citep{2013ARA&A..51..511K} launch collimated bipolar relativistic jets perpendicular to the accretion disk \citep{1977MNRAS.179..433B, 1982MNRAS.199..883B}. These jets inflate lobes that propagate far into the surrounding medium, interacting with both the circumgalactic and intergalactic environments. Understanding the evolution of these systems, including the duty cycle of radio activity and the role of the environment in jet propagation, is crucial to understanding the feedback in the formation of large-scale structures \citep{2008ApJ...677...63S, 2015MNRAS.449..955M, 2021MNRAS.502.5104L, 2024A&A...686A..21S, 2025A&A...693A..77G}.

The physical conditions that allow GRGs to reach such extraordinary sizes are still not fully understood. Several hypotheses have been proposed to explain it. One scenario suggests that they are very old sources whose jets have had enough time to expand into the Intergalactic Medium (IGM) \citep{1998A&A...329..431M, 2004A&A...421..899L}. Alternatively, they may host particularly powerful or well-collimated jets, or they may inhabit underdense environments that offer minimal resistance to jet propagation \citep{2008MNRAS.385.1286J, 2015MNRAS.449..955M}. However, recent work challenges the environmental explanation, showing that a significant fraction of GRGs, $\sim$10$\%$, in some surveys are located in galaxy clusters, where the ambient density is expected to be high \citep{2020yCat..36350005D, 2020MNRAS.499...68T, 2021MNRAS.502.5104L}. This suggests that factors beyond the local environment, such as intrinsic nuclear properties or episodic jet activity, may also be critical in driving GRG evolution.

An increasingly supported scenario is that many GRGs are relics of past AGN activity, now experiencing a second phase of accretion and jet launching, called a restarted AGN \citep{2012ApJS..199...27S, 2017MNRAS.471.3806K, 2020MNRAS.494..902B, 2023Galax..11...74M}. Evidence for such restarting include double-double morphologies, where new inner lobes are embedded within older outer lobes, as well as discrepancies between the current nuclear luminosity and the power inferred from the radio lobes \citep{2018MNRAS.481.4250U}. Hard X-ray observations, in particular, have shown that the nuclei of many GRGs are remain active, emitting strongly in the $\sim$10 – 100 keV band, even when their radio lobes appear aged and disconnected from the core \citep{2016MNRAS.461.3165B, 2018MNRAS.481.4250U, 2019MNRAS.486.3975K, 2020MNRAS.494..902B, 2021A&A...650A..51M, 2021MNRAS.503.4681B}. Also, see a comprehensive review of GRGs by \cite{2023JApA...44...13D}, with a particular focus on their high-energy emission properties.

Despite their size and significance, the nuclear properties of GRGs have remained poorly explored in X-rays, especially at soft energies. In this work, we present a systematic X-ray spectral study of 27 GRGs selected from \textit{XMM-Newton} and \textit{Chandra} archival observations. We perform detailed spectral modeling to characterize the nature and geometry of the circumnuclear material, the presence of reprocessing features associated with cold reflection, and the potential impact of warm, ionized gas in the nuclear regions. We also explore how the nuclear X-ray luminosity compares with the extended radio power, as a potential diagnostic of the evolutionary stage or re-triggered jet activity. By connecting spectral features to physical processes in the AGN core, this study aims to provide new constraints on the accretion and feedback mechanisms operating in evolved, large-scale radio galaxies.

This paper is organized as follows. In Section~\ref{data}, we describe the sample selection, X-ray data and radio data reduction. In Section~\ref{Analysis}, we outline the spectral modeling methodology. The results of the spectral analysis are presented in Section~\ref{spectral properties}, followed by a discussion in Section~\ref{discussion}. Finally, we summarize our conclusions in Section~\ref{conclusion}.

%%%%%%%%%%\newpage

%\clearpage
%\newpage
\section{DATA}
\label{data}

\subsection{Sample selection}

The sources chosen for the analysis here are drawn from the catalogue of GRGs by \cite{2017MNRAS.469.2886D}, \cite{2020yCat..36350005D}, \cite{2020A&A...642A.153D} and \cite{2018ApJS..238....9K}. We analysed $\sim$756 GRGs detected in the LOFAR Two-metre Sky Survey (LoTSS) at 144 MHz, the NRAO VLA Sky Survey (NVSS) at 1400 MHz, and the Sydney University Molonglo Sky Survey (SUMSS) at 843 MHz. We cross-correlated these 756 GRGs with the \textit{XMM-Newton} DR14 (4XMM) catalogue within a search radius of 15$^{\prime\prime}$ and the \textit{Chandra} Source Catalog (CSC; version 2.1) within a search radius of 1$^{\prime\prime}$ \citep{2019yCat.9057....0E} to identify GRGs observed in the soft X-ray band. The different search radii account for the different angular resolutions of the two X-ray observatories. This cross-matching identified 54 GRGs in 4XMM and 13 in CSC, yielding a final sample of 67 GRGs (approximately 9\% of the parent sample), listed in Table~\ref{Appendix Table A1}.

We note that this selection introduces an observational bias toward relatively X-ray bright, and consequently preferentially nearby GRGs. As a result, the sample is not statistically complete or representative of the entire GRG population. For the present work, we further restricted the sample to sources with sufficiently high signal-to-noise ratio (S/N) to enable reliable X-ray spectral fitting and robust determination of the spectral parameters. The conclusions presented in this work should therefore be interpreted as applicable to the X-ray detected subset of GRGs rather than to the full GRG population.

\subsection{Chandra analysis}
The archival data for the 13 GRGs observed by \textit{Chandra} were uniformly reprocessed from the level 1 event file with CIAO 4.17.0 and CALDB 4.11.6, following standard techniques detailed in the online documentation of the CIAO.\footnote{http://asc.harvard.edu/ciao/} The reduction pipeline included reprocessing with \texttt{chandra\textunderscore{repro}}, filtering for high background periods, and removing bad pixels. 

We inspected the data using an astronomical imaging and data visualization application SAOImageDS9 and defined a circular source region centred on each target with a radius of $\sim$30 arcsec. For background subtraction, we selected a nearby source-free region of similar radius. Spectra and calibration files (ARF and RMF) were extracted using the script \texttt{specextract}.

Among the 13 extracted spectra, only 6 exhibited an acceptable S/N ($>$20). In the remaining 7 spectra, the background contribution dominated too early, or the number of data points was too few to constrain the spectra. Therefore, we restricted our further analysis to these 6 sources from CSC.

\begin{table*}
\small
\caption{List of GRGs selected for analysis in X-ray band, ordered by redshift.}
\centering
\begin{adjustbox}{width=0.88\textwidth}
\begin{tabular}{@{\extracolsep{14pt}}ccccccc}
\hline
No. & IAU Name & Obs. ID & RA (h:m:s) & DEC (d:m:s) & $z$ & Size (Mpc)\\ \hline
1 & 4XMM J005748.8+302109 & 0305290201 & 00h 57m 48.838s & +30° 21' 09.01" & 0.0164 & 1.18 \\
2 & 4XMM J010724.9+322445 & 0305290101 & 01h 07m 24.991s & +32° 24' 45.60" & 0.0167 & 1.13 \\
3 & 4XMM J040845.9-624751 & 0881390101 & 04h 08m 54.911s & -62° 47' 51.27" & 0.0178 & 0.71 \\
4 & 4XMM J044909.0+450039 & 0146490101 & 04h 49m 09.057s & +45° 00' 39.33" & 0.0208 & 0.76 \\
5 & 4XMM J163232.1+823216 & 0056340201 & 16h 32m 32.145s & +82° 32' 16.68" & 0.0240 & 1.97 \\
6 & 4XMM J111125.3+265748 & 0892420201 & 11h 11m 25.340s & +26° 57' 48.93" & 0.0335 & 1.12 \\
7 & 4XMM J074837.0+554900 & 0404050101 & 07h 48m 37.004s & +55° 49' 00.67" & 0.0356 & 1.21 \\
8 & 4XMM J131217.0+445021 & 0892420301 & 13h 12m 17.020s & +44° 50' 21.99" & 0.0358 & 0.99 \\
9 & 4XMM J050549.2-283519 & 0892420401 & 05h 05m 49.335s & -28° 35' 20.08" & 0.0381 & 1.87 \\
10 & 4XMM J233355.2-234340 & 0760990201 & 23h 33m 55.226s & -23° 43' 40.37" & 0.0477 & 1.09 \\
11 & 4XMM J162804.0+514631 & 0500850501 & 16h 28m 04.067s & +51° 46' 31.59" & 0.0547 & 1.25 \\
12 & 4XMM J094945.9+731423 & 0404050601 & 09h 49m 45.921s & +73° 14' 23.17" & 0.0581 & 1.02 \\
13 & 2CXO J201801.2-553931 & 0000032177 & 20h 18m 01.270s & -55° 39' 31.60" & 0.0606 & 1.37 \\
14 & 4XMM J011202.2+492834 & 0655610101 & 01h 12m 02.208s & +49° 28' 34.91" & 0.0677 & 0.88 \\
15 & 4XMM J111305.6+401730 & 0406610501 & 11h 13m 05.630s & +40° 17' 30.45" & 0.0745 & 1.05 \\
16 & 4XMM J234532.7-044924 & 0800630101 & 23h 45m 32.717s & -04° 49' 24.83" & 0.0755 & 1.41 \\
17 & 2CXO J155209.2+200523 & 0000010908 & 15h 52m 09.202s & +20° 05' 23.09" & 0.0895 & 2.04 \\
18 & 4XMM J031819.1+682932 & 0312190501 & 03h 18m 19.174s & +68° 29' 32.37" & 0.0901 & 1.54 \\
19 & 2CXO J100601.7+345410 & 0000010249 & 10h 06m 01.738s & +34° 54' 10.28" & 0.1005 & 4.58 \\
20 & 4XMM J235904.3-605459 & 0677180601 & 23h 59m 04.305s & -60° 54' 59.05" & 0.1014 & 0.75 \\
21 & 4XMM J204237.3+750802 & 0200910201 & 20h 42m 37.306s & +75° 08' 02.43" & 0.1040 & 1.18 \\
22 & 4XMM J051555.7-805941 & 0604530201 & 05h 15m 55.721s & -80° 59' 41.55" & 0.1050 & 0.89 \\
23 & 4XMM J003405.4-663935 & 0653880301 & 00h 34m 05.499s & -66° 39' 35.34" & 0.1070 & 1.93 \\
24 & 4XMM J001119.6+321709 & 0827021201 & 00h 11m 19.690s & +32° 17' 09.78" & 0.1071 & 0.95 \\
25 & 4XMM J124733.3+672316 & 0306680201 & 12h 47m 33.392s & +67° 23' 16.19" & 0.1072 & 1.50 \\
26 & 4XMM J154858.0-321657 & 0601260301 & 15h 48m 58.063s & -32° 16' 57.55" & 0.1085 & 1.02 \\
27 & 2CXO J045252.8+520447 & 0000014957 & 04h 52m 24.426s & +52° 04' 48.83" & 0.1090 & 1.15 \\
28 & 4XMM J104755.1+741935 & 0147450301 & 10h 47m 55.106s & +74° 19' 35.76" & 0.1210 & 1.51 \\
29 & 4XMM J144851.0-400846 & 0720280101 & 14h 48m 51.032s & -40° 08' 46.14" & 0.1230 & 1.54 \\
30 & 4XMM J142932.9+544342 & 0871191201 & 14h 29m 32.908s & +54° 43' 42.85" & 0.1232 & 1.07 \\
31 & 4XMM J031301.9+412001 & 0306680301 & 03h 13m 01.957s & +41° 20' 01.18" & 0.1340 & 1.40 \\
32 & 4XMM J093952.7+355358 & 0852580101 & 09h 39m 52.759s & +35° 53' 58.95" & 0.1367 & 0.78 \\
33 & 4XMM J020354.4-041358 & 0677610139 & 02h 03m 54.493s & -04° 13' 58.82" & 0.1375 & 1.14 \\
34 & 4XMM J140948.5-030236 & 0694510101 & 14h 09m 48.555s & -03° 02' 36.41" & 0.1376 & 1.53 \\
35 & 4XMM J214529.1+815454 & 0403000101 & 21h 45m 29.135s & +81° 54' 54.21" & 0.1457 & 2.89 \\
36 & 4XMM J011625.0-472241 & 0601260101 & 01h 16m 25.036s & -47° 22' 41.29" & 0.1461 & 1.85 \\
37 & 2CXO J132834.1-012917 & 0000013981 & 13h 28m 34.157s & -01° 29' 17.10" & 0.1510 & 0.87 \\
38 & 2CXO J010944.3+731157 & 0000009295 & 01h 09m 44.334s & +73° 11' 57.10" & 0.1810 & 0.75 \\
39 & 4XMM J172320.7+341758 & 0102040101 & 17h 23m 20.744s & +34° 17' 58.33" & 0.2060 & 0.79 \\
40 & 4XMM J013929.8+395712 & 0671640201 & 01h 39m 29.856s & +39° 57' 12.61" & 0.2107 & 1.21 \\
41 & 4XMM J225336.0-345530 & 0882810101 & 22h 53m 36.033s & -34° 55' 36.03" & 0.2115 & 0.96 \\
42 & 2CXO J143103.6+334541 & 0000018595 & 14h 31m 03.610s & +33° 45' 41.28" & 0.2383 & 1.03 \\
43 & 4XMM J174838.9-233520 & 0720280201 & 17h 48m 38.987s & -23° 35' 20.29" & 0.2400 & 1.31 \\
44 & 4XMM J092438.2+302837 & 0553440601 & 09h 24m 38.270s & +30° 28' 37.19" & 0.2730 & 0.82 \\
45 & 4XMM J133603.8+030732 & 0601781201 & 13h 36m 03.844s & +03° 07' 32.34" & 0.3213 & 0.79 \\
46 & 4XMM J142735.6+263214 & 0841481501 & 14h 27m 35.603s & +26° 32' 14.48" & 0.3636 & 1.05 \\
47 & 2CXO J142105.6+414448 & 0000010381 & 14h 21m 05.649s & +41° 44' 48.42" & 0.3670 & 0.97 \\
48 & 4XMM J123526.6+212034 & 0671640801 & 12h 35m 26.769s & +21° 20' 34.45" & 0.4220 & 0.87 \\
49 & 4XMM J231207.6+184541 & 0502500101 & 23h 12m 07.615s & +18° 45' 41.99" & 0.4265 & 1.17 \\
50 & 4XMM J143046.4+145012 & 0862181501 & 14h 30m 46.426s & +14° 50' 12.38" & 0.4351 & 1.59 \\
51 & 4XMM J013528.4+375405 & 0600450501 & 01h 35m 28.466s & +37° 54' 05.20" & 0.4373 & 0.94 \\
52 & 2CXO J151100.0+075150 & 0000013886 & 15h 11m 00.029s & +07° 51' 50.07" & 0.4593 & 0.77 \\
53 & 2CXO J042925.8+003304 & 0000021392 & 04h 29m 25.850s & +00° 33' 04.85" & 0.4680 & 1.90 \\
54 & 2CXO J114720.7-125309 & 0000010414 & 11h 47m 20.737s & -12° 53' 09.67" & 0.4966 & 0.92 \\
55 & 4XMM J152311.0+520303 & 0840440201 & 15h 23m 11.080s & +52° 03' 03.50" & 0.5166 & 0.72 \\
56 & 4XMM J085743.5+394528 & 0551630601 & 08h 57m 43.552s & +39° 45' 28.82" & 0.5288 & 1.05 \\
57 & 2CXO J125142.0+503424 & 0000016062 & 12h 51m 42.030s & +50° 34' 24.76" & 0.5490 & 0.87 \\
58 & 4XMM J121952.3+472058 & 0400560301 & 12h 19m 52.327s & +47° 20' 58.24" & 0.6535 & 1.60 \\
59 & 4XMM J135041.7+642936 & 0147540101 & 13h 50m 41.792s & +64° 29' 36.57" & 0.7100 & 0.98 \\
60 & 2CXO J145307.9+221707 & 0000021438 & 14h 53m 07.900s & +22° 17' 07.71" & 0.7848 & 0.89 \\
61 & 4XMM J133127.8+250049 & 0401260101 & 13h 31m 27.844s & +25° 00' 50.00" & 0.8040 & 0.77 \\
62 & 4XMM J020448.3-094409 & 0763910301 & 02h 04m 48.316s & -09° 44' 09.59" & 1.0033 & 2.08 \\
63 & 4XMM J221944.2-202130 & 0556180201 & 22h 19m 44.284s & -20° 21' 30.07" & 1.1480 & 0.73 \\
64 & 4XMM J010424.2-660924 & 0653880101 & 01h 04m 24.239s & -66° 24' 24.12" & 1.1900 & 0.75 \\
65 & 4XMM J021659.2-044945 & 0112371701 & 02h 16m 58.915s & -04° 49' 23.67" & 1.3250 & 1.24 \\
66 & 4XMM J235522.9+795518 & 0556180101 & 23h 55m 22.976s & +79° 55' 18.07" & 1.3360 & 0.71 \\
67 & 4XMM J090816.7+394325 & 0400830301 & 09h 08m 16.744s & +39° 43' 25.94" & 1.8830 & 0.95 \\
\hline
\end{tabular}
\label{Appendix Table A1}
\end{adjustbox}
\end{table*}

\subsection{XMM-Newton analysis}
For \textit{XMM-Newton} observations, pre-processed spectra and light curves were initially retrieved from the \textit{XMM-Newton Science Archive (XSA)}\footnote{https://nxsa.esac.esa.int/} for sources with publicly available pipeline-processed data. For sources lacking archived spectra, manual extraction was performed using the Science Analysis System (SAS v1.3), following the workflow described in the \textit{XMM-Newton} Data Processing Workshop \citep{2023eas..conf.1030Y}. This process included filtering high background intervals with \texttt{evselect} and \texttt{tabgtigen}, defining source and background regions, and generating calibrated response files with \texttt{rmfgen} and \texttt{arfgen}. However, several pre-processed and manually extracted spectra exhibited low S/N, likely due to short exposure times or faint source fluxes. These spectra were excluded from spectral analysis.   

Among the 54 sources observed by \textit{XMM-Newton}, only 21 exhibited an acceptable S/N of more than 20. Therefore, for our final spectral analysis, we restricted ourselves to these 21 sources from 4XMM and 6 sources from CSC (see Table~\ref{Table 1} - \ref{Table 6}).

\begin{figure}
    \centering
    \includegraphics[width=1.0\linewidth]{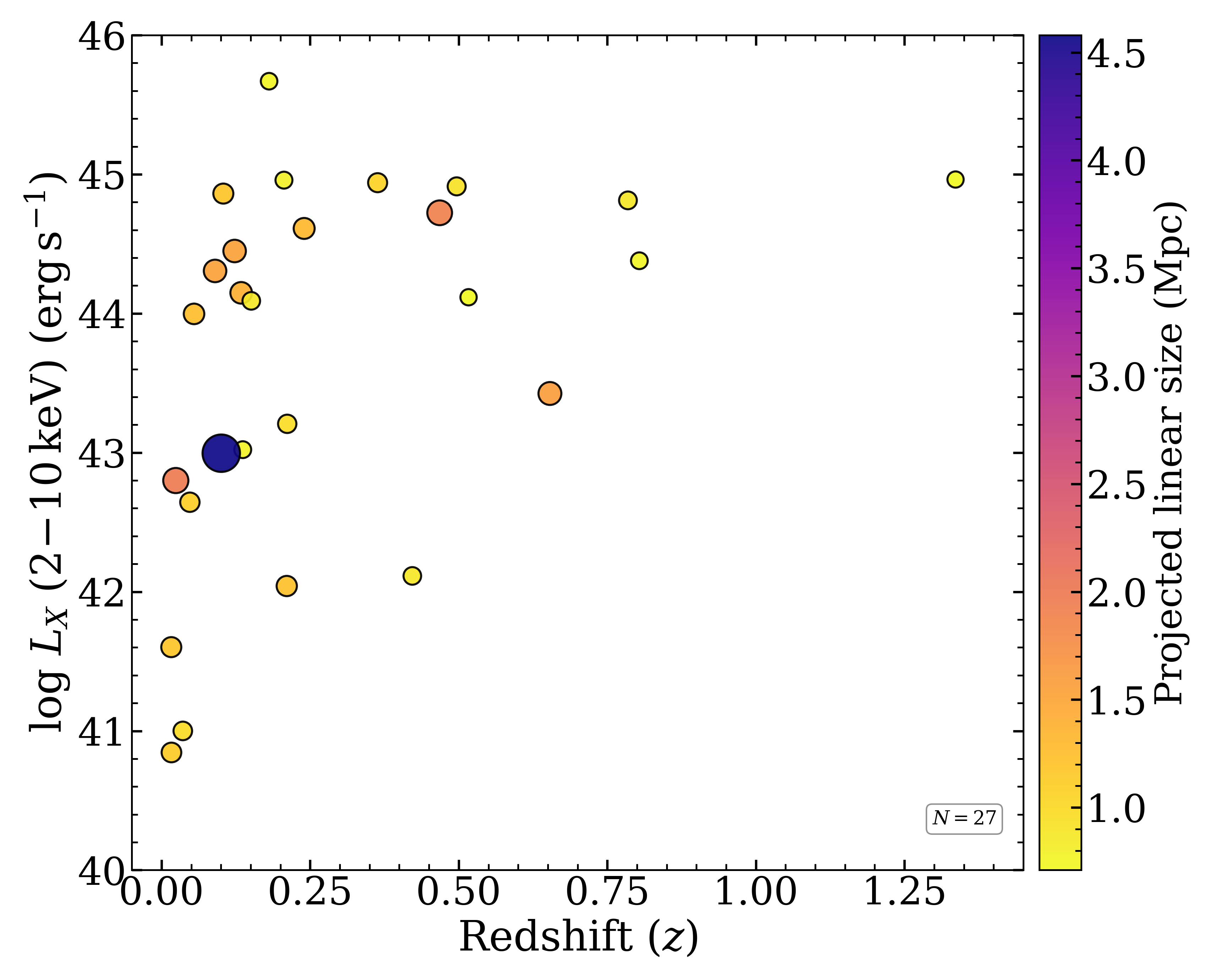}
    \caption{X-ray flux distribution with redshift for the 27 GRGs.}
    \label{fig: Luminosity vs Redshift}
\end{figure}

To assess whether the final spectroscopic sample of 27 GRGs is a statistically fair representation of the X-ray cross-matched sample from which it was drawn, we compared the redshift and projected linear-size distributions of the 27 analysed sources (see Fig.~\ref{fig: Luminosity vs Redshift}) with those of the full 67 source cross-matched sample listed in Table~\ref{Appendix Table A1} (i.e.\ the 54 4XMM and 13 CSC matches, prior to the additional S/N$>$20 cut applied for spectral fitting). The 27 analysed GRGs span $z = 0.016$--$1.336$ (median $z = 0.18$) and linear sizes of $0.71$--$4.58$~Mpc (median $1.09$~Mpc), essentially indistinguishable from the full 67 source sample ($z$ median $= 0.14$, range $0.016$--$1.883$; size median $= 1.05$~Mpc, range $0.71$--$4.58$~Mpc). A two-sample Kolmogorov--Smirnov (KS) test finds no significant difference between the analysed (27) and excluded (40, low S/N) subsamples in either redshift ($D = 0.16$, $p = 0.76$) or linear size ($D = 0.10$, $p = 0.99$), and likewise no significant difference between the analysed subsample and the full 67 source cross-matched sample ($D = 0.09$, $p = 0.99$ for $z$; $D = 0.06$, $p = 1.00$ for size). This indicates that the S/N based selection applied in this work does not introduce an additional, statistically significant redshift or a size bias. We caution, however, that this test only probes bias introduced at the second selection step (the S/N cut on the X-ray spectra). The first step (cross-matching the parent catalogue of $\sim$756 GRGs against the 4XMM and CSC catalogs) is itself expected to preferentially select X-ray bright, and hence relatively nearby and/or radio core luminous, GRGs, as already noted above in the sample selection.

\subsection{Radio Analysis}

We performed a radio morphological and spectral analysis of our sample of GRGs to identify and quantify the flux contributions from their cores and lobes. Using TGSS-NVSS\footnote{https://tgssadr.strw.leidenuniv.nl/doku.php?id=spidx} spectral index maps \citep{2018MNRAS.474.5008D}, we visually identified the core and extended lobe regions. These maps allowed us to distinguish flat-spectrum core emission from steep-spectrum lobe emission. For each source, we defined regions around the core and lobes, and extracted integrated flux densities with CASA-VIEWER \citep{2022PASP..134k4501C} using the \texttt{imstat} tasks. For some GRGs in our sample, the radio core emission could not be clearly separated from the surrounding diffuse emission. In such cases, we measured only the lobe flux densities. By following the method described in \cite{2018MNRAS.481.4250U}. The corresponding radio powers were calculated using the standard luminosity-distance relation by \cite{2010ApJ...720.1066C}:

\begin{equation}
P_\nu = 4 \pi D_L^2 \, S_\nu \, (1 + z)^{\alpha - 1} ,
\end{equation}

where \( D_L \) is the luminosity distance, \( S_\nu \) is the measured flux density at frequency \( \nu \), \( z \) is the redshift, and \( \alpha \) is the radio spectral index. We assumed that the spectra behave as $S_{\nu} \propto \nu^{-\alpha}$, and adopted a fixed value of $\alpha$ = 0.7 for the spectral index for all GRGs in our sample, consistent with typical values reported in previous studies of GRGs \citep{2020A&A...642A.153D}.

\section{Spectral Analysis}
\label{Analysis}
Spectral analysis for all 27 sources was performed using \texttt{XSPEC} v.12.12.0, in the 0.3-10 keV energy range, with the fitting range varying based on background dominance. 

\subsection*{Model 1: Galactic and intrinsic absorption with a power-law} 
Initially, we attempted to fit the spectra with a single power-law along with fixed Galactic absorption \texttt{phabs*zpo}. This model provided a good fit for 10 spectra. For other 5 spectra, we identified the need for additional intrinsic absorption at the source redshift, which was modelled using  \texttt{phabs*zwabs*zpo} in \texttt{XSPEC}. These two models resulted in good fits for a total of 15 source spectra (see Table~\ref{Table 1}), though in three cases, low degrees of freedom (dof) (e.g. 7, 9, 10) led to weaker spectral constraints. Of these 15 spectra, 12 showed no evidence for an Fe K$\alpha$ emission line. For the remaining three sources (2CXO J100601.7+345410, 4XMM J133127.8+250049, 4XMM J031301.9+412001), our extended search for the line (see Sect.~\ref{spectral properties}) yielded a significant detection; for these we added a narrow Gaussian component with the centroid energy fixed at the rest-frame value of 6.40~keV and width fixed at 10~eV, resulting in the equivalent widths and spectral fits parameters are reported in Table~\ref{Table 1}.

\subsection*{Model 2: Absorption Edges and Emission Lines:}
Although our baseline model provided good fits for many sources, some spectra exhibited residuals suggesting additional components. In particular, some sources showed absorption edges, probably from ionised material, while others also displayed a Fe K$\alpha$ emission line around 6.4 keV. To account for these characteristics, we used the \texttt{phabs*zwabs*zedge*(zpo + zga)} model, incorporating one or two absorption edges along with a Gaussian line to the fit. This model improved the fits for 3 of our sources where absorption edges were apparent at ~0.7–2.7 keV, likely associated with O, Ne, or Fe transitions. Additionally, in two of these spectra of 4XMM J174838.9-233520 and 4XMM J204237.3+750802 sources, we detected the Fe K$\alpha$ emission line with varying equivalent widths. The result of this model is summarised in Table~\ref{Table 2}.

\subsection*{Model 3: Modeling Soft Excess Emission with $\texttt{swind1}$}

In some spectra, we observed a soft excess below 2 keV. We initially explored standard thermal models, such as \texttt{diskbb} and \texttt{apec}, to  describe this component. While these models provided satisfactory fits for some sources (see Model 4), they failed to adequately reproduce the soft excess in a subset of spectra, often yielding statistically poor fits or physically implausible parameter values.

For these sources, we considered the possibility that the soft excess might actually be due to smeared absorption from partially ionised material rather than an intrinsic emission component \citep{2007ASPC..373..121D}. A previous study by \cite{2015MNRAS.451.2370M} found that the \texttt{swind1} model provided a good fit for one of the source 4XMM J144851.0-400846 considered in our analysis, so we applied the same approach to our data. While it has been noted that this model can predict unrealistically high outflow velocities \citep{2008MNRAS.386L...1S}, it remains a useful approximation for a partially covering, ionised absorber, which could explain the observed soft excess.

The final model \texttt{phabs*zwabs*zedge*swind1*(zpo+zga)}, provided improved fits for 3 sources exhibiting significant soft excess below $\sim$2 keV. A Fe K$\alpha$ line was already detected for two of these sources 4XMM J144851.0-400846 and 4XMM J142735.6+263214. For the third source (4XMM J163232.1+823216), the line was not originally detected; however, our extended search for the line (see Sect.~\ref{spectral properties}) yielded a significant detection, with the line centroid left free to vary ($E=6.49^{+0.07}_{-0.08}$~keV) and its width fixed at 10~eV, giving the equivalent width and the result of this model is summarised in Table~\ref{Table 3}.

\subsection*{Model 4: Modeling Soft Excess Emission with \texttt{apec}}
For 4 sources shown in Table~\ref{Table 4}, we found that the soft excess could be better explained by a thermal plasma emission model rather than smeared absorption. In these cases, we used a combination of a power-law component (\texttt{zpo}) to represent non-thermal emission and a thermal plasma component (\texttt{apec}) to model the excess emission. The final  model \texttt{phabs*zwabs*zedge*(zpo+apec)}, provided a significantly improved fit for these spectra. However, the abundance parameter in the \texttt{apec} model remained unconstrained, likely due to the limited signal-to-noise ratio and the absence of strong metal emission lines. To stabilise the fit, we fixed the abundance to solar metallicity (Z = 1 \Zsun), which is a commonly adopted assumption in cases where the metallicity cannot be reliably determined from the data. None of these four sources showed a formally detected Fe K$\alpha$ line in the original analysis. However, our extended search for the line (see Sect.~\ref{spectral properties}) yielded a significant detection for source 4XMM J172320.7+341758; for this source, a narrow Gaussian component was added to the model (\texttt{phabs*zwabs*zedge*(zpo+apec+zga)}), with the centroid energy fixed at the rest-frame value of 6.40~keV and width fixed at 10~eV, giving the equivalent width and the result is summarised in Table~\ref{Table 4}.

\subsection*{Model 5: Partial covering and Fe K$\alpha$ emission}

For one source 4XMM J162804.0+514631 showing moderate absorption and hints of soft scattered emission or weak reflection features, we employed a model: \texttt{phabs*(zpo + zwabs*zpo + zga)}. This model combines Galactic absorption (\texttt{phabs}), an intrinsic absorbed power-law (\texttt{zwabs*zpo}) representing the obscured AGN continuum, an additional unabsorbed power-law (\texttt{zpo}) to account for soft scattered or jet-related emission, and a Gaussian line (\texttt{zga}) modelling potential Fe K$\alpha$ emission near 6.4 keV. The result is summarised in Table \ref{Table 5}.

\subsection*{Model 6: Broken Power-law Model}
4XMM J093952.7+355358 source best fitted with a Broken Power-law model (\texttt{bknpower}) to account for a change in spectral slope, combined with a narrow Gaussian component ($\texttt{zgauss}$) to model the Fe K$\alpha$ emission line. The model describes the continuum with two photon indices, $\Gamma_1$ and $\Gamma_2$, below and above the break energy $E_{\rm break}$. The result is summarised in Table~\ref{Table 6}.

%\newpage
%\clearpage
\onecolumn
\begin{landscape}
\begin{longtable}[c]{cccccccccc}
      \caption{Model 1: Galactic and Intrinsic absorption with a power-law. Note : $^{\dagger}$ Double -Double Radio Galaxy (DDRG). $^{f}$ - fixed the parameter. }
\\ \hline
No.& Source & Type  & $N_{H}$ (galactic)& $N_{H}$ (Intrinsic) & $\Gamma$ & $E$ & $\sigma$ & $EW$ & $\chi^{2}/{dof}$ \\
&  &    & ($10^{22}$) & ($10^{22}$) &  & \\
&  &    &  (cm$^{-2}$)        &     (cm$^{-2}$)  &  &(keV) & (eV) &  (eV) &\\
\hline
\vspace{5pt}
54 & 2CXO J114720.7-125309 & FR II  & 0.029 & - -  & 1.58$^{+0.06}_{-0.06}$ & - -  & - -  & - - & 162/185 \\
\vspace{5pt}
19 &2CXO J100601.7+345410 & DDRG$^{\dagger}$  & 0.010 & 1.96$^{+0.43}_{-0.38}$ & 1.52$^{+0.25}_{-0.23}$ & 6.40$^{f}$ & 10$^{f}$ & 224.63$^{+257.07}_{-218.67}$ & 98/121 \\
\vspace{5pt}
53 & 2CXO J042925.8+003304 & - -   & 0.059 & 4.82$^{+0.86}_{-0.78}$ & 1.49$^{+0.18}_{-0.17}$ &  - -  & - -  & - - & 127/136 \\
\vspace{5pt}
60 & 2CXO J145307.9+221707 & FR II  & 0.026 & - -  & 1.24$^{+0.15}_{-0.14}$ & - -  & - -  & - - & 38/38 \\
\vspace{5pt}
37 & 2CXO J132834.1-012917 & FR II  & 0.026 & - -  & 1.18$^{+0.06}_{-0.06}$ &  - -  & - -  & - - & 140/162 \\
\vspace{5pt}
38 & 2CXO J010944.3+731157 & FR II  & 0.206 & 5.21$^{f}$ & 1.35$^{+0.12}_{-0.12}$ & - -  & - -  & - - & 150/143 \\
\vspace{5pt}
58 & 4XMM J121952.3+472058 & FR II  & 0.026 & - -  & 1.78$^{+0.07}_{-0.07}$ & - -  & - -  & - - & 35/40  \\
\vspace{5pt}
61 & 4XMM J133127.8+250049 & FR II  & 0.010 & 0.13$^{+0.08}_{-0.07}$ & 1.4$^{f}$ & 6.40$^{f}$ & 10$^{f}$ & $391.56^{+387.03}_{-318.90}$ & 30/28 \\
\vspace{5pt}
66 & 4XMM J235522.9+795518 & FR II  & 0.129 & 10.72$^{+4.19}_{-3.37}$ & 1.29$^{+0.26}_{-0.24}$ & - -  & - -  & - - & 8/10 \\
\vspace{5pt}
40 & 4XMM J013929.8+395712 & FR I  & 0.055 & - -  & 2.16$^{+0.25}_{-0.24}$ &  - -  & - -  & - - & 5/7   \\
\vspace{5pt}
48 & 4XMM J123526.6+212034 & FR II  & 0.024 & - -  & 1.52$^{+0.22}_{-0.22}$ & - -  & - -  & - - & 8/9  \\
\vspace{5pt}
10 & 4XMM J233355.2-234340 & FR II  & 0.017 & - -  & 1.74$^{+0.01}_{-0.01}$ & - -  & - -  & - - & 155/163 \\
\vspace{5pt}
55 & 4XMM J152311.0+520303 & FR II  & 0.015 & - -  & 1.75$^{+0.06}_{-0.06}$ & - -  & - -  & - - & 41/42  \\
\vspace{5pt}
41 & 4XMM J225336.0-345530 & FR II  & 0.010 & - -  & 1.84$^{+0.03}_{-0.03}$ & - -  & - -  & - - & 73/63  \\
\vspace{5pt}
31 & 4XMM J031301.9+412001 & FR I  & 0.101 & - -  & 1.75$^{+0.02}_{-0.02}$ & 6.40$^{f}$ & 10$^{f}$ & 151.69$^{+83.53}_{-70.98}$ & 104/121 \\
\hline
\label{Table 1}
\end{longtable}

\small
\begin{longtable}[c]{ccccccccccccc}
\caption{Model 2: Absorption Edges and Emission lines}      
\\ \hline
No. & Source & Type  & $N_{H}$ (Intrinsic) & $\Gamma$ & $E$ & $\tau$ & $E$ & $\tau$ & $E$ & $\sigma$ & $EW$ & $\chi^{2}/{dof}$ \\
 &  &  &  ($10^{22}$) & & & & & & & & &  \\
 &  &  &  (cm$^{-2}$)     & & (keV) &  & (keV) &   & (keV) &  (eV) & (eV) & \\
\hline
\vspace{5pt}
18 & 4XMM J031819.1+682932 & FR II  & 3.78$^{+0.26}_{-0.25}$ & 1.87$^{+0.05}_{-0.05}$ & 2.22$^{+0.05}_{-0.05}$ & 0.49$^{+0.12}_{-0.12}$ & - -  &  - - & - - & - - & - - & 113/109 \\ 
\vspace{5pt}
43 & 4XMM J174838.9-233520 & FR II  & 1.89$^{+0.17}_{-0.17}$ & 1.62$^{+0.08}_{-0.08}$ & 2.72$^{+0.05}_{-0.05}$ & 0.33$^{+0.10}_{-0.10}$ & 1.59$^{+0.03}_{-0.03}$ & 0.75$^{+0.16}_{-0.15}$ & 6.52$^{+0.09}_{-0.09}$ & 251.25$^{+98.62}_{-86.30}$ & 200$^{+67.53}_{-54.42}$ & 109/110 \\
\vspace{5pt}
21 & 4XMM J204237.3+750802 & FR II  & 0.029$^{+0.005}_{-0.005}$ & 1.69$^{+0.01}_{-0.01}$ & 0.74$^{+0.01}_{-0.01}$ & 0.23$^{+0.04}_{-0.04}$ & 0.91$^{+0.02}_{-0.01}$ & 0.12$^{+0.03}_{-0.03}$ & 6.44$^{+0.03}_{-0.03}$ & 121.78$^{+43.45}_{-36.16}$ & 118.53$^{+23.01}_{-21.63}$ & 197/158  \\

\hline
\label{Table 2}
\end{longtable}
\end{landscape}

\onecolumn
\begin{landscape}
{\small
\setlength{\tabcolsep}{3pt} % default is 6pt

\begin{longtable}[c]{cccccccccccccccc}
\caption{Model 3: Modeling Soft Excess Emission with \texttt{swind1}. Note : $^{f}$ - fixed the parameter.}      
\\ \hline
No. & Source & Type  & $N_{H}$ (Intrinsic) & $\Gamma$ & $E$ & $\tau$ & $E$ & $\tau$ & $N_{H}$ (swind1) & $Log(\xi)$ & $\sigma$ & $E$ & $\sigma$ & $EW$ & $\chi^{2}/dof$  \\
&  & & ($10^{22}$) & &  & & & & ($10^{22}$) &  & \\
&  & & (cm$^{-2}$) & & (keV) & & (keV) & &  (cm$^{-2}$) & (erg cm$^{-1}$ s$^{-1}$) & & (keV) & (eV) & (eV) &  \\
\hline
\vspace{5pt}
5 & 4XMM J163232.1+823216 & - -  & 0.108$^{+0.01}_{-0.01}$ & 2.12$^{+0.03}_{-0.03}$ & 0.48$^{+0.02}_{-0.02}$ & 0.23$^{+0.06}_{0.06}$ & - - & - - & 10.60$^{+1.66}_{-1.49}$ & 2.93$^{+0.09}_{-0.08}$ & 0.42$^{f}$ & 6.49$^{+0.07}_{-0.08}$ & 10$^{f}$ & 61.29$^{+31.07}_{-40.04}$ & 132/136 \\

\vspace{5pt}
29 & 4XMM J144851.0-400846 & FR II  & 0.215$^{+0.03}_{-0.03}$ & 1.87$^{+0.09}_{-0.08}$ & 0.70$^{+0.005}_{-0.006}$ & 2.86$^{+0.29}_{-0.27}$ & 0.92$^{+0.03}_{-0.03}$ & 0.77$^{+0.20}_{-0.19}$ & 12.54$^{+2.99}_{-2.25}$ & 2.52$^{+0.08}_{-0.07}$ & 0.36$^{f}$ & 6.45$^{+0.03}_{-0.03}$ & 30$^{f}$ & 85.69$^{+25.86}_{15.90}$  & 154/151 \\

\vspace{5pt}
46 & 4XMM J142735.6+263214 & FR II  & - - & 2.02$^{+0.01}_{-0.01}$ & 0.76$^{+0.01}_{-0.01}$ & 0.52$^{+0.03}_{-0.04}$ & - - & - - & 12.81$^{+1.64}_{-1.47}$ & 3.03$^{+0.05}_{-0.05}$ & 0.35$^{+0.04}_{-0.03}$ & 6.33$^{+0.04}_{-0.04}$ & 20$^{f}$ & 63.98$^{+21.11}_{15.39}$ & 167/156 \\
\hline
\label{Table 3}
\end{longtable}
}

\begin{longtable}[c]{ccccccccccccccccccc}
\caption{Model 4: Modeling Soft Excess Emission with \texttt{apec}. Note : $^{f}$ - fixed the parameter. }      
\\ \hline
No. & Source & Type  & $N_{H}$ (Intrinsic) & $\Gamma$ & $E$ & $\tau$ & KT (\texttt{apec}) & $E$ & $\sigma$ & $EW$ & $\chi^{2}/dof$  \\
&  & & ($10^{22}$) & &  & & &   &  & & \\
&  & & (cm$^{-2}$) & & (keV) & & (keV) & (keV) & (eV) & (eV) &  \\
\hline
\vspace{5pt}
2 & 4XMM J010724.9+322445 & FR I  & - - & 2.01$^{+0.08}_{-0.08}$ & - - & - - &  0.80$^{+0.03}_{-0.03}$ & - -  & - -  & - - & 40/42 \\
\vspace{5pt}
39 & 4XMM J172320.7+341758 & FR II  & - - & 1.92$^{+0.02}_{-0.02}$ & - - & - - & 0.19$^{+0.01}_{-0.01}$ & 6.40$^{f}$ & 10$^{f}$ & 151.01$^{+68.89}_{-61.40}$ & 113/110  \\
\vspace{5pt}
1 & 4XMM J005748.8+302109 & - -  & 0.10$^{+0.02}_{-0.02}$ & 1.69$^{+0.09}_{-0.08}$ & 1.07$^{+0.01}_{-0.02}$ & 0.41$^{+0.08}_{-0.08}$ & 0.67$^{+0.01}_{-0.01}$ & - -  & - -  & - - & 92/85 \\
8 & 4XMM J131217.0+445021 & FR II  & - - & 1.5$^{f}$ & - - & - - & 0.67$^{+0.03}_{-0.04}$ & - -  & - -  & - -  & 24/20 \\
\hline
\label{Table 4}
\end{longtable}

\begin{longtable}[c]{cccccccccccccccccc}
\caption{Model 5: Partial covering and Fe K$\alpha$ emission}      
\\ \hline
No. & Source & Type  & $N_{H}$ (Intrinsic) & $\Gamma(abs.)$ & $\Gamma(unabs)$ & $E$ & $\sigma$  & $EW$ &$\chi^{2}/dof$  \\
 & & & ($10^{22}$) & &  & & &   &  & & \\
 & & & (cm$^{-2}$) & &  & (keV)  & (eV) & (eV) &  &  \\
 \hline
 \vspace{5pt}
11 & 4XMM J162804.0+514631 & FR II  & 16.9$^{+0.6}_{-0.8}$ & 1.65$^{+0.16}_{-0.15}$ & 2.80$^{+0.15}_{-0.14}$ & 6.39$^{+0.07}_{-0.07}$ &  207$^{+95.81}_{-92.88}$ & 218.48$^{+75.88}_{-71.15}$ & 118/116 \\

\hline
\label{Table 5}
\end{longtable}

\begin{longtable}[c]{cccccccccccccccc}
\caption{Model 6:  \texttt{Broken power-law} model }      
\\ \hline
No. & Source & Type  & $\Gamma$$_{1}$ & $\Gamma$$_{2}$ & $E_{break}$ & $E$ & $\sigma$  & EW &$\chi^{2}/dof$  \\
  & & &  & & &   &  & & \\
  & & &  & & (keV) & (keV) & (eV) & (eV)  &  \\
 \hline
 \vspace{5pt}
32 & 4XMM J093952.7+355358 & FR II   & 2.01$^{+0.30}_{-0.24}$ & 0.39$^{+0.15}_{-0.16}$ & 1.45$^{+0.32}_{-0.27}$ & 6.44$^{+0.05}_{-0.05}$ &  20$^{f}$ & 567.66$^{+179.67}_{-223.16}$ & 23/23 \\

\hline
\label{Table 6}
\end{longtable}
\end{landscape}

\twocolumn

\section{Spectral Properties}
\label{spectral properties}

In our sample of 27 GRGs, 12 sources show moderate to significant intrinsic absorption ($N_\mathrm{H,int} \gtrsim 10^{21}$ cm$^{-2}$), while the remainder are adequately fit with only Galactic absorption. This suggests that most GRGs in our study host unobscured or only lightly obscured nuclei. The low levels of obscuration may result from favorable orientation, consistent with the AGN unification scheme \citep{1993ARA&A..31..473A, 1995PASP..107..803U}, or may reflect physical clearing of circumnuclear gas by jet-driven feedback processes \citep{2008NewAR..52..227T, 2005A&A...439..521M}. Statistical studies have shown that compact radio sources typically exhibit stronger X-ray absorption than extended ones \citep{2002ApJ...568..592B}, supporting an evolutionary scenario in which expanding jets disrupt or disperse surrounding material.

To place the obscuration properties of our GRG sample in the broader context of the AGN population, we compared our $N_{\rm H}$ distribution with published samples of radio-loud and radio-quiet AGN. Among the 12/27 ($\sim$44\%) GRGs with detected intrinsic absorption, the column densities span $N_{\rm H,int} \sim (0.03$--$17)\times10^{22}$~cm$^{-2}$, with a median of $\sim1.9\times10^{22}$~cm$^{-2}$. Considering the full sample of 27 sources (i.e.\ treating non-detections as unabsorbed), 7/27 ($\sim$26\%) exhibit $N_{\rm H}>10^{22}$~cm$^{-2}$. This fraction is statistically consistent (Fisher's test, $p=0.24$) with the $\sim$40\% absorbed fraction reported by \citet{2016MNRAS.461.3153P} for a hard X-ray selected sample of 64 radio galaxies drawn from the INTEGRAL/IBIS and Swift/BAT surveys. It is also intermediate between the unabsorbed radio-quiet Type~1 and the heavily absorbed Type~2 populations of the \citet{2011A&A...530A..42C} XBS sample, for which the type dichotomy is defined at $N_{\rm H}=4\times10^{21}$~cm$^{-2}$. This level of obscuration is also broadly consistent with the ASCA study of \citet{1999ApJ...526...60S}, who found that broad-line radio galaxies (BLRGs, the optical analogues of Type~1 unification) are typically less absorbed than narrow-line radio galaxies and radio galaxies (Type~2 analogues), the latter reaching $N_{\rm H}\gtrsim10^{22}$--$10^{23}$~cm$^{-2}$ in a substantial fraction of cases. \citet{2016MNRAS.461.3153P} further report that $\sim$75\% of narrow-line hard X-ray selected radio galaxies are absorbed above this threshold, versus a much smaller fraction among broad-line objects, in agreement with the unification picture. Our GRGs populate an intermediate regime between these extremes, consistent with a mix of viewing angles and/or genuinely reduced covering factors of circumnuclear gas, as might be expected for radio galaxies whose extended jets have already begun to clear the nuclear environment \citep{2008NewAR..52..227T, 2002ApJ...568..592B}.

Among the 12 GRGs with intrinsic absorption, several exhibit additional spectral complexity. For example, 4XMM J162804.0+514631 source requires both a high column density ($N_\mathrm{H} \sim 10^{23}$ cm$^{-2}$) and a partial covering component, consistent with a clumpy or inhomogeneous absorber. This model is in line with contemporary AGN torus frameworks, where the obscuring medium consists of a distribution of dusty, optically thick clouds with variable covering fractions \citep{2013ApJ...771...87H, 2012AdAst2012E..17B, 2017NatAs...1..679R}. Such clumpy geometries naturally explain the simultaneous presence of heavily absorbed primary emission and a weaker, scattered component.

Narrow Fe K$\alpha$ emission lines centered near 6.4–6.5 keV are detected in 6 of the GRGs showing intrinsic absorption, with equivalent widths (EWs) ranging from $\sim$60 eV to over 500 eV. These features originate from fluorescence in cold, optically thick gas, likely in the torus or broad-line region, and indicate X-ray reflection from circumnuclear structures \citep{1994MNRAS.267..743G, 2014MNRAS.441.3622R}. For instance, 4XMM J162804.0+514631 source shows a broad line with EW $\sim$218 eV, while source 4XMM J093952.7+355358 exhibits a particularly strong line (EW $\sim$568 eV). These reflection features are predominantly seen in sources modeled with partial covering or smeared absorption (e.g., 4XMM J142735.6+263214, 4XMM J144851.0-400846), but are absent in sources modeled solely with thermal plasma (\texttt{apec}), consistent with the Fe K$\alpha$ origin in compact nuclear regions rather than diffuse hot gas. The presence of strong, narrow Fe K$\alpha$ lines confirms that GRG nuclei retain classical reflection structures similar to those seen in Seyfert galaxies.

Several GRGs also exhibit absorption edges in the soft X-ray regime (0.7–2.7 keV), likely arising from partially ionized warm absorbers near the nucleus. These features are often attributed to outflowing gas or disk winds associated with AGN activity \citep{2003ARA&A..41..117C, 2005A&A...431..111B}. A soft X-ray excess below $\sim$2 keV is also detected in multiple sources. In some cases, this component is best modeled with a smeared ionized absorber (\texttt{swind1}), indicating high-velocity winds \citep{2007ASPC..373..121D, 2015MNRAS.451.2370M}, while in others, a thermal plasma model (\texttt{apec}) provides a better fit, consistent with emission from hot (kT $\sim 0.2$–0.8 keV) gas in the host galaxy or surrounding group \citep{2008MNRAS.386.1709C}.

To assess the statistical significance of the absorption edges and \texttt{swind1} components introduced in Models~2 and~3, we performed an $F$-test comparing each adopted model with the corresponding baseline model \texttt{phabs*zwabs*zpo} for the same source (Table~\ref{Table Ftest}). The additional component(s) improve the fit by $\Delta\chi^{2}=18$--$202$ for 2--7 additional free parameters, corresponding to $F$-values of 6.9--47.2 and $F$-test probabilities $p_F$ ranging from $3\times10^{-4}$ down to $<10^{-15}$. The edge and/or \texttt{swind1} component(s) are therefore required at very high confidence in all six sources, confirming that these features are statistically robust rather than artefacts of overfitting. The absorption edge energies ($E\simeq0.7$--$2.7$~keV) are consistent with O~{\sc vii}/O~{\sc viii} and Ne/Fe-L transitions, the same lines responsible for warm absorber signatures in roughly half of Seyfert~1 galaxies \citep{1997MNRAS.286..513R,2026A&A...710A.118M}. For the three \texttt{swind1} fitted sources we derive $N_{\rm H}\simeq(1.1$-$1.3)\times10^{23}\,\mathrm{cm}^{-2}$ and $\log\xi\simeq2.5$--$3.0$, somewhat higher in column density and at the upper end in ionization compared with classical Seyfert warm absorbers \citep[$N_{\rm H}\sim10^{20}$--$10^{22}\,\mathrm{cm}^{-2}$, $\log\xi\sim0$--$2$;][]{2005A&A...431..111B,2007MNRAS.379.1359M} and with warm absorbers previously reported in broad-line radio galaxies \citep[$\log\xi\simeq2$--$3$;][]{2012MNRAS.419..321T}, placing our sources in a transition regime between classical warm absorbers and ultra-fast outflows \citep{2010A&A...521A..57T}. We caution that the \texttt{swind1} outflow velocities can be unreliable \citep{2008MNRAS.386L...1S}, and therefore base this comparison on the better constrained column density and ionization parameter alone.

In 4XMM J093952.7+355358 a broken power-law fit reveals a soft X-ray slope consistent with coronal emission below 1.45 keV, and a significantly hardened continuum above the break. This hardening, coupled with a strong Fe K$\alpha$ line, suggests the presence of a jet-dominated hard X-ray component possibly accompanied by reflection features. The break at $\sim$1.45 keV marks the transition between thermal and non-thermal dominance in the spectrum, typical of low-accretion rate, radio-loud AGNs.

Taken together, these results reveal a rich diversity in the circumnuclear environments of GRGs. While most nuclei are unobscured, a significant subset exhibits spectral features consistent with clumpy absorbers, reflection off cold material, warm ionized outflows, and extended thermal gas. This diversity likely reflects a combination of orientation effects, AGN feedback mechanisms (particularly jet-driven clearing), and evolutionary processes associated with the long lifespans and large scales of GRG jets \citep{2008NewAR..52..227T, 2002ApJ...568..592B, 2020NewAR..8801539H}.

To more fully characterise the Fe~K$\alpha$ statistics of our sample, we extended our search for the line to the 21 sources that originally lacked a formal detection, by adding a narrow ($\sigma$ fixed at 10~eV) Gaussian component fixed at rest-frame 6.4~keV to the best-fit continuum model of each source. This procedure yielded 5 additional significant detections, bringing the total number of Fe~K$\alpha$ detections in the sample to 11 (Table~\ref{Table FeK}). For 8 of the remaining 16 sources we determined a meaningful 90\% single-sided confidence upper limit on the EW via $\Delta\chi^{2}=2.7$; for the remaining 8 sources, the spectral quality was insufficient to place a useful constraint, and these are marked with a dash in Table~\ref{Table FeK}. The 11 detected EWs (median $\simeq$~152~eV, mean $\simeq$~203~eV, range $\sim$61--568~eV) are broadly consistent with previous measurements for broad-line radio galaxies, although our sample extends to somewhat higher values. Earlier \textit{ASCA} and \textit{RXTE} studies reported typical EWs of $\sim$90--150~eV \citep{1999ApJ...526...60S,2000ApJ...537..654E}. These values are systematically weaker than the $\sim$100--200~eV typically observed in radio-quiet Seyfert~1 galaxies, a difference commonly attributed to dilution of the reflection signature by beamed, jet-linked X-ray emission in radio-loud sources. We further tested for a possible anti-correlation between EW and $L_{2-10\,\mathrm{keV}}$ among the 11 detected sources (the Iwasawa--Taniguchi effect; \citealt{1993ApJ...413L..15I}), finding $\log(EW/\mathrm{eV}) = (-0.09\pm0.12)\,\log L_{2-10\,\mathrm{keV}} + 6.4$ (Pearson $r=-0.25$, $p=0.46$; Spearman $\rho=-0.31$, $p=0.36$). We do not find a statistically significant anti-correlation between Fe~K$\alpha$ EW and X-ray luminosity in our sample. The absence of a significant Iwasawa--Taniguchi effect in the full sample may reflect genuine physical differences between our GRG population and the radio-quiet/Seyfert samples in which the effect was originally identified. For example, dilution of the reprocessed reflection component by jet-linked emission, or a narrower dynamic range in $L_{2-10\,\mathrm{keV}}$ than probed in larger AGN samples -- as well as the still modest sample size and the exclusion of the 16 non-detections (with upper limits or unconstrained EWs) from this test. A survival analysis (censored) correlation test incorporating the upper limits, and application to a larger GRG sample, would help clarify whether a weaker version of this effect is present.

\section{Discussion}
\label{discussion}

\subsection{Bolometric Luminosity}

To investigate how the nuclear and large-scale jet power relate to GRGs, we compared the bolometric luminosity derived from X-ray emission with that inferred from the extended radio lobes. Our analysis closely follows the methodology adopted by \cite{2018MNRAS.481.4250U}, including the use of the \cite{2011MNRAS.417L..51V} correlation to estimate the bolometric luminosity from the 1.4 GHz radio lobe luminosity, and the bolometric correction of \citet{2004MNRAS.351..169M} to estimate the bolometric luminosity from X-rays. This allows for a direct comparison between the time-averaged jet power traced by the extended radio lobes and the nuclear activity reflected in the X-ray emission.

\begin{figure}
    \centering
    \includegraphics[width=1.0\linewidth]{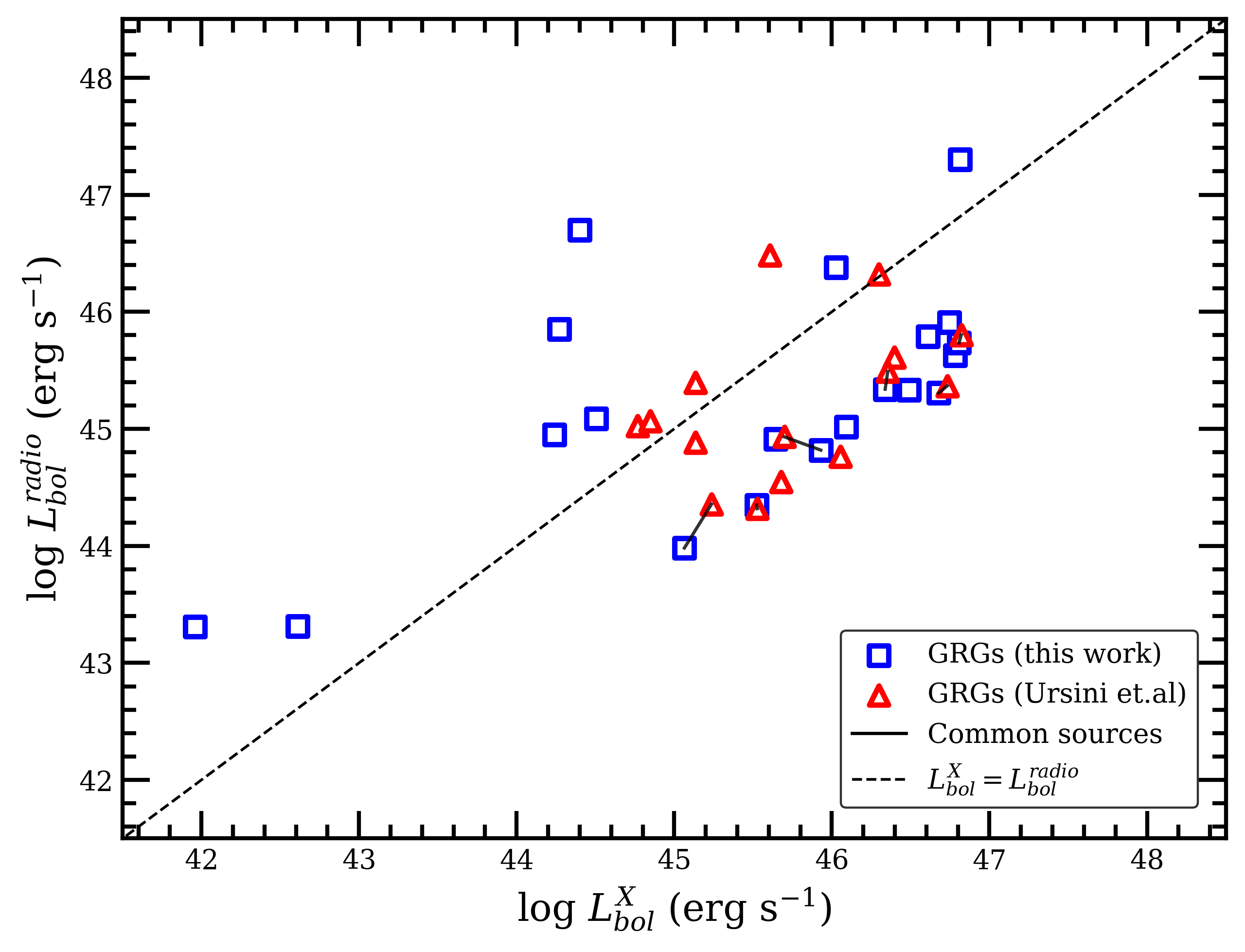}
    \caption{Bolometric luminosity estimated from the radio luminosity of the lobes (y-axis) versus that estimated from the 2-10 keV luminosity (x-axis). Blue squares denote the GRGs of our sample, overlaid in the plot of \protect\cite{2018MNRAS.481.4250U}. The dashed line represents the identity y=x. Common sources present in both samples are connected by solid black line.}
    \label{fig:Lbol_vs_Rbol}
\end{figure}

The radio-based bolometric luminosity, $L_{\mathrm{bol}}^{\mathrm{radio}}$, was calculated using the \cite{2011MNRAS.417L..51V} relation:

\begin{table*}
    \centering
    \caption{X-ray and radio fluxes and luminosities: column (3) 2-10 keV flux; column (4) 2-10 keV luminosity; column (5) core flux density at 1.4 GHz; column (6) lobes flux density at 1.4 GHz; column (7) core luminosity at 1.4 GHz; and column (8) lobes luminosity at 1.4 GHz. For some GRGs (*) in our sample, the radio core emission could not be clearly separated from the surrounding diffuse emission. In such cases, we measured only the lobe flux densities.}
     \begin{tabular}{cccccccc}
\\ \hline
No. & IAU Name & $F$ (2-10 keV) & $L$ (2-10 keV) & $S_{\mathrm{1.4\,GHz}}^{\mathrm{core}}$ & $S_{\mathrm{1.4\,GHZ}}^{\mathrm{lobes}}$ & $L_{\mathrm{1.4\,GHz}}^{\mathrm{core}}$ & $L_{\mathrm{1.4\,GHz}}^{\mathrm{lobes}}$ \\
& &  ($10^{-13}$ erg $\mathrm{s}^{-1}$ $\mathrm{cm}^{-2}$) & ($10^{42}$ erg $\mathrm{s}^{-1}$) & (mJy) &  (mJy) & ($10^{40}$ erg $\mathrm{s}^{-1}$) & ($10^{40}$ erg $\mathrm{s}^{-1}$) \\
\hline
21 & 4XMM J204237.3+750802 & 270.6 & 726.8 & 212.8 & 1562 & 7.4 & 54.4 \\

43 & 4XMM J174838.9-233520 & 25.3 & 408.3 & 61.02 & 327.1 & 10.9 & 58.7 \\

18 & 4XMM J031819.1+682932 & 100.3 & 202.1 & 76.4 & 675.6 & 2.0 & 17.7  \\

2 & 4XMM J010724.9+322445 & 1.2 & 0.07 & -- & --  & --   & --    \\

46 & 4XMM J142735.6+263214 & 18.9 & 870.4 & 72.6 & 286.5 & 28.8 & 113.6  \\

11 & 4XMM J162804.0+514631 & 141.9 & 99.3 & 41.3 & 617.3 & 0.4 & 6 \\

5 & 4XMM J163232.1+823216 & 47.9 & 6.3 & -- & -- &  --  &  --   \\

48 & 4XMM J123526.6+212034$^{*}$ & 0.2 & 1.3 & --   & 2563 & --   & 1344 \\

61 & 4XMM J133127.8+250049$^{*}$ & 1.1 & 239.2 & --  & 389.2 & --   & 645.1 \\

1 & 4XMM J005748.8+302109$^{*}$ & 7.1 & 0.4 & -- & 618.3 & --   & 0.6 \\

39 & 4XMM J172320.7+341758 & 74.3 & 907.8 & 510.5 & 1094 & 68.1 & 145.8  \\

10 & 4XMM J233355.2-234340 & 83.5 & 4.4 & 791 & 348.1 & 5.9 & 2.6 \\

58 & 4XMM J121952.3+472058 & 1.6 & 26.6 & 8.3 & 24.3 & 9.6 & 28.1 \\

41 & 4XMM J225336.0-345530 & 1.3 & 16.1 & 45.8 & 234.2 & 6.4 & 32.9 \\

31 & 4XMM J031301.9+412001 & 30.3 & 140.7 & --  &  --   & --   &  --    \\

8 & 4XMM J131217.0+445021 & 0.4 & 0.1 & 151.1 & 130.6 & 0.6 & 0.5 \\

40 & 4XMM J013929.8+395712 & 0.08 & 1.1 & --   &  -- &  --   & --   \\

55 & 4XMM J152311.0+520303 & 1.4 & 130.8 & --   & --  &  --   &  --  \\

66 & 4XMM J235522.9+795518$^{*}$ & 1.5 & 917.4 &  --  & 1444 &  --   & 5386 \\

32 & 4XMM J093952.7+355358 & 2.6 & 10.5 & 161.5 & 3252 & 9.7 & 192.2 \\

29 & 4XMM J144851.0-400846 & 71.8 & 281.1 & --   &  --  &  --   &  --  \\

38 & 2CXO J010944.3+731157 & 30.2 & 4666 &  --  &  --  &  --   &  --  \\

19 & 2CXO J100601.7+345410 & 4.1 & 9.9 & 3269 & 734.7 & 106.5 & 23.9 \\

60 & 2CXO J145307.9+221707$^{*}$ & 3.4 & 649.6 & -- & 104.1 &  --  & 165.6 \\

54 & 2CXO J114720.7-125309$^{*}$ & 10.1 & 819.2 & --  & 310.2 & --   & 219.6 \\

37 & 2CXO J132834.1-012917$^{*}$ & 22.2 & 123.2 &  --  & 303 &  --   & 22 \\

53 & 2CXO J042925.8+003304 & 7.7 & 528.9 & 20.7 & 90.8 & 13.1 & 57.7 \\
\hline

\end{tabular}
\label{Table 8}
\end{table*}

\begin{equation}
\log \left( \frac{L_{1.4\,\mathrm{GHz}}^{\mathrm{lobes}}}{L_{\mathrm{bol}}} \right) = -3.57,
\end{equation}
with an associated scatter of 0.47 dex. For the X-ray based bolometric luminosity, $L_{\mathrm{bol}}^X$, we adopted the luminosity-dependent bolometric correction of \citet{2004MNRAS.351..169M} applied to the 2--10 keV X-ray luminosities. While the intrinsic scatter in this relation is relatively small ($\sim$0.1 dex), the dominant uncertainty arises from source variability \citep{2009MNRAS.392.1124V}. By using the same calibrations and assumptions as \cite{2018MNRAS.481.4250U}, we aim to assess whether the bolometric behaviour observed in their FR~II radio galaxies extends to our sample of GRGs.

In Fig.~\ref{fig:Lbol_vs_Rbol}, we overplot our GRG sample (blue squares) on top of the \cite{2018MNRAS.481.4250U} (red triangles). Most of our GRG sources follow a similar trend: $L_{\mathrm{bol}}^X$ is generally higher than $L_{\mathrm{bol}}^{\mathrm{radio}}$, often by nearly an order of magnitude, especially at high luminosities ($\log L_{\mathrm{bol}}^X \gtrsim 44.5$). This result supports the interpretation that X-rays trace the current accretion state, while radio lobes represent a long-term average of past jet activity. A discrepancy between the two can thus arise naturally if AGNs undergo intermittent accretion episodes \citep[e.g.][]{2013A&A...557L...7V, 2018MNRAS.481.4250U}.

Interestingly, a few GRGs (4XMM J131217.0+445021, 4XMM J005748.8+302109, 4XMM J123526.6+212034, 4XMM J225336.0-345530, 4XMM J093952.7+355358, 2CXO J100601.7+345410) in our sample, particularly at lower X-ray luminosities exhibit the opposite trend: radio-derived bolometric luminosities exceeding the X-ray-based estimates. Although these points are few and may be affected by small-number statistics, they may suggest an evolutionary scenario where the central engine is currently in a low accretion state, while the lobes still reflect a prior high-luminosity phase. This may indicate a delayed radiative response in the lobes or ageing synchrotron plasma, phenomena expected in GRGs due to their extended lifetimes.

Alternatively, the discrepancy could point to limitations in the application of the \cite{2011MNRAS.417L..51V} relation to GRGs. The relation was originally calibrated on smaller, optically selected FR-II sources and may not fully capture the environmental, structural or evolutionary differences present in GRGs. Radiative losses, inverse compton cooling, or jet composition (e.g. electron-positron pairs) may alter the observed synchrotron luminosity for a given jet power, leading to an underestimate of $L_{\mathrm{bol}}^{\mathrm{radio}}$.

Despite these uncertainties, the general behaviour in our GRG sample aligns with the trend reported by \cite{2018MNRAS.481.4250U}, reinforcing the broader picture that X-ray and radio luminosities, while linked, probe different timescales and phases of AGN activity. Our results suggest that GRGs may serve as powerful laboratories for studying this decoupling, especially at late evolutionary stages where fossil lobes coexist with fading or retriggered nuclei.

\subsection{Radio and X-ray core luminosity correlation}
\label{sec:core_radio}

In addition to the extended-lobe radio powers used above, Table~\ref{Table 8} also lists the 1.4~GHz core flux densities and derived core luminosities ($L_{\mathrm{1.4\,GHz}}^{\mathrm{core}}$) for the subset of GRGs in which the compact core could be cleanly separated from the diffuse lobe emission in the TGSS--NVSS spectral-index maps. Reliable core measurements are available for 13 of the 27 sources; the remaining 14 either lack a detectable core or show core emission blended with the surrounding diffuse lobe flux (marked with an asterisk in Table~\ref{Table 8}).

The core luminosities in our sample span $\log(L_{\mathrm{1.4\,GHz}}^{\mathrm{core}}/\mathrm{erg\,s^{-1}}) \simeq 39.6$--$42$ (median $\simeq 41$), roughly 2 to 3 orders of magnitude below the corresponding 2-10 keV nuclear luminosities ($\log L_{2-10\,\mathrm{keV}} \simeq 41$--$45$, median $\simeq 44$). This is qualitatively consistent with the core luminosities reported for other GRG samples selected in hard X-rays \citep[e.g.][]{2018MNRAS.481.4250U, 2020MNRAS.494..902B, 2021MNRAS.503.4681B, 2016MNRAS.461.3165B}, which similarly find $L_{2-10\,\mathrm{keV}}$ to dominate over $L_{\mathrm{1.4\,GHz}}^{\mathrm{core}}$ by several orders of magnitude for radiatively efficient nuclei.

\begin{figure}
    \centering
    \includegraphics[width=1.0\linewidth]{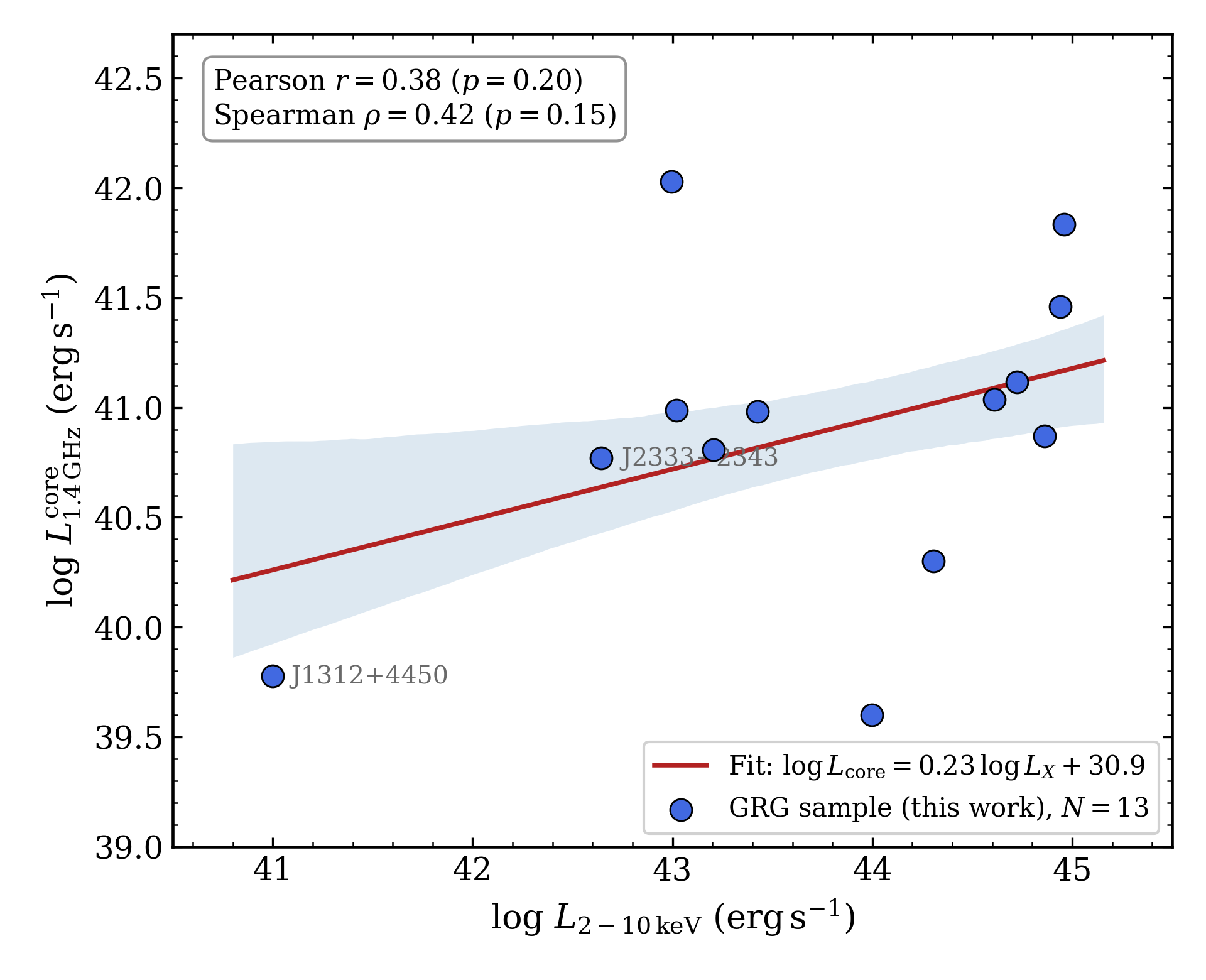}
    \caption{Radio core luminosity ($L_{1.4\,\mathrm{GHz}}^{\mathrm{core}}$) versus 2--10~keV nuclear X-ray luminosity for the 13 GRGs in our sample with a resolvable radio core. The red line shows the best-fit linear relation in log--log space, with the shaded band indicating the 1$\sigma$ bootstrap uncertainty on the fit; Pearson and Spearman coefficients are reported in the inset. The two lowest $L_X$ sources discussed in the text (4XMM~J131217.0+445021 and 4XMM~J233355.2$-$234340) are labelled.}
    \label{fig:LX_Rcore}
\end{figure}

We tested a correlation between the nuclear X-ray luminosity and the radio core luminosity for the 13 sources with measured cores (see Fig.~\ref{fig:LX_Rcore}). We find a positive but statistically weak trend (Pearson $r = 0.38$, $p = 0.20$; Spearman $\rho = 0.42$, $p = 0.15$), with a best-fit relation $\log L_{\mathrm{1.4\,GHz}}^{\mathrm{core}} = (0.23 \pm 0.15)\,\log L_{2-10\,\mathrm{keV}} + (30.9 \pm 6.5)$. The correlation is considerably weaker and not statistically significant compared with the tight $L_X$--$L_{\mathrm{core}}$ correlation found by \cite{2018MNRAS.481.4250U} for their hard X-ray selected GRG sample (later confirmed with a larger sample by \cite{2021MNRAS.500.3111B}, who report a Kendall's $\tau = 0.44$, $p = 5\times10^{-3}$). We attribute this difference primarily to (i) the much smaller number of GRGs in our sample with resolvable radio cores ($N=13$, versus $N\gtrsim20$ in the hard X-ray-selected studies), (ii) the fact that our sample is selected in the soft (0.3-10 keV) band rather than hard X-rays, which should be less biased toward the most radiatively efficient, core-dominated nuclei, and (iii) non-simultaneity between the archival radio (TGSS/NVSS) and X-ray (\textit{XMM-Newton}/\textit{Chandra}) observations, for variable AGN cores can create a difference in the correlation.

Despite the weak statistical significance of the correlation, the observed trend broadly suggests that higher nuclear X-ray luminosities are generally associated with higher radio core luminosities. This is consistent with a scenario in which the compact radio core and the X-ray emitting corona (or jet base) are physically linked, with both tracing the instantaneous accretion state of the central engine \citep{2018MNRAS.481.4250U}. We note that two of the GRGs with the lowest $L_{2-10\,\mathrm{keV}}$ in our resolvable core (4XMM J131217.0+445021 and 4XMM J233355.2$-$234340) also have relatively high $L_{\mathrm{1.4\,GHz}}^{\mathrm{core}}/L_{2-10\,\mathrm{keV}}$ ratios, potentially indicating either heavily obscured, X-ray faint nuclei or radio cores dominated by an unresolved, compact jet component rather than a simple corona. A larger sample with simultaneous radio and X-ray coverage, and reliable core/lobe deblending for all 27 sources, will be needed to establish whether GRGs selected in soft X-rays follow the same $L_X$--$L_{\mathrm{core}}$ relation established for hard X-ray selected GRGs.

\begin{figure}
    \centering
    \includegraphics[width=1.0\linewidth]{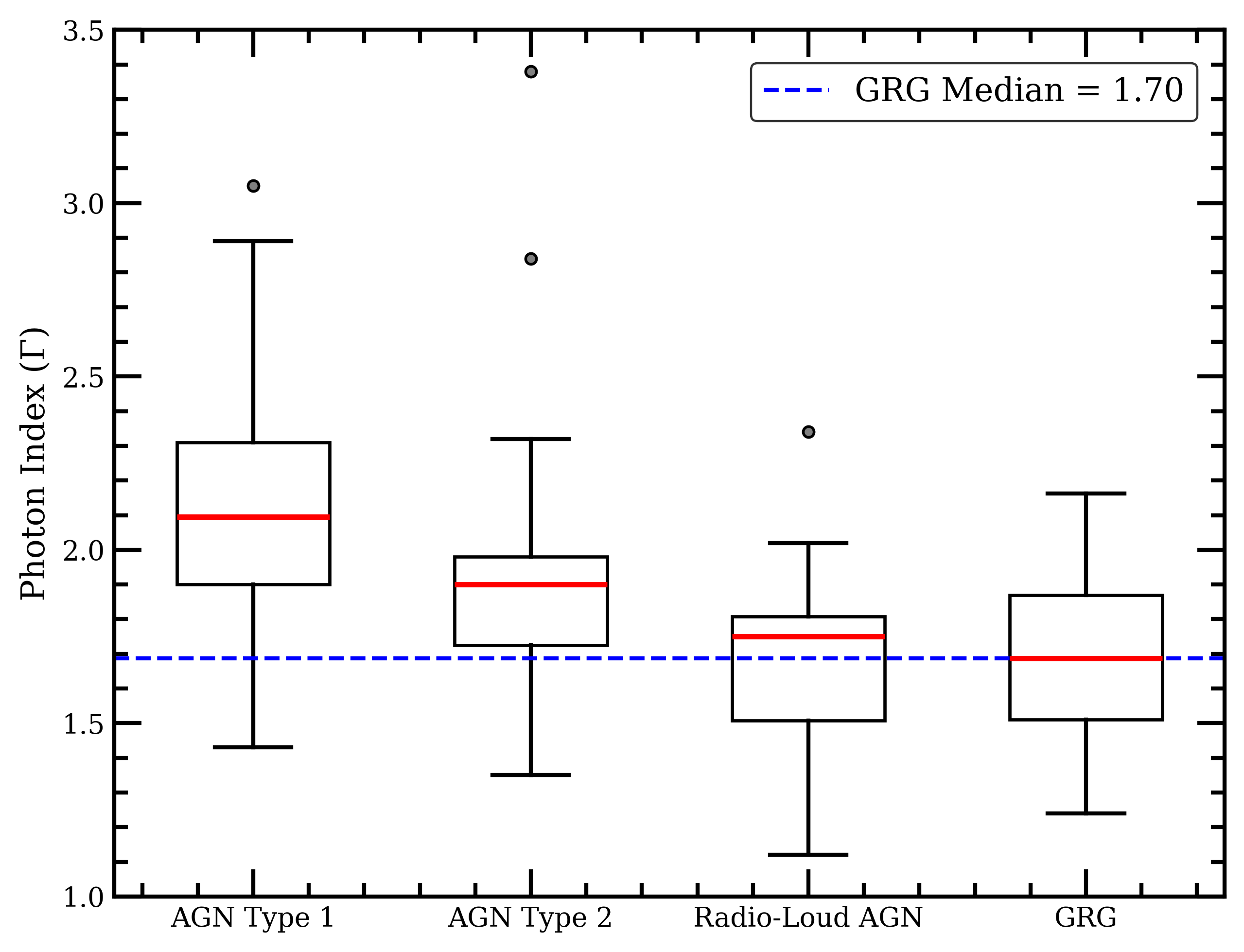}
    \caption{Box plot comparing the X-ray photon index ($\Gamma$) distributions across four AGN classes: Radio-quite Type 1 AGNs \& Type 2 AGNs, radio-loud AGNs from \citet{2011A&A...530A..42C,1999ApJ...526...60S}, and our GRG sample. The red lines indicate the median $\Gamma$ for each class, while the blue dashed line marks the median $\Gamma$ value of the GRG sample ($\Gamma$ = 1.70). The plot shows that GRGs and radio-loud AGNs generally have harder X-ray spectra (lower $\Gamma$) compared to radio-quiet Type 1 and Type 2 AGNs.}
    \label{fig:enter-photon}
\end{figure}

\begin{figure}
\centering
\includegraphics[width=1.0\linewidth]{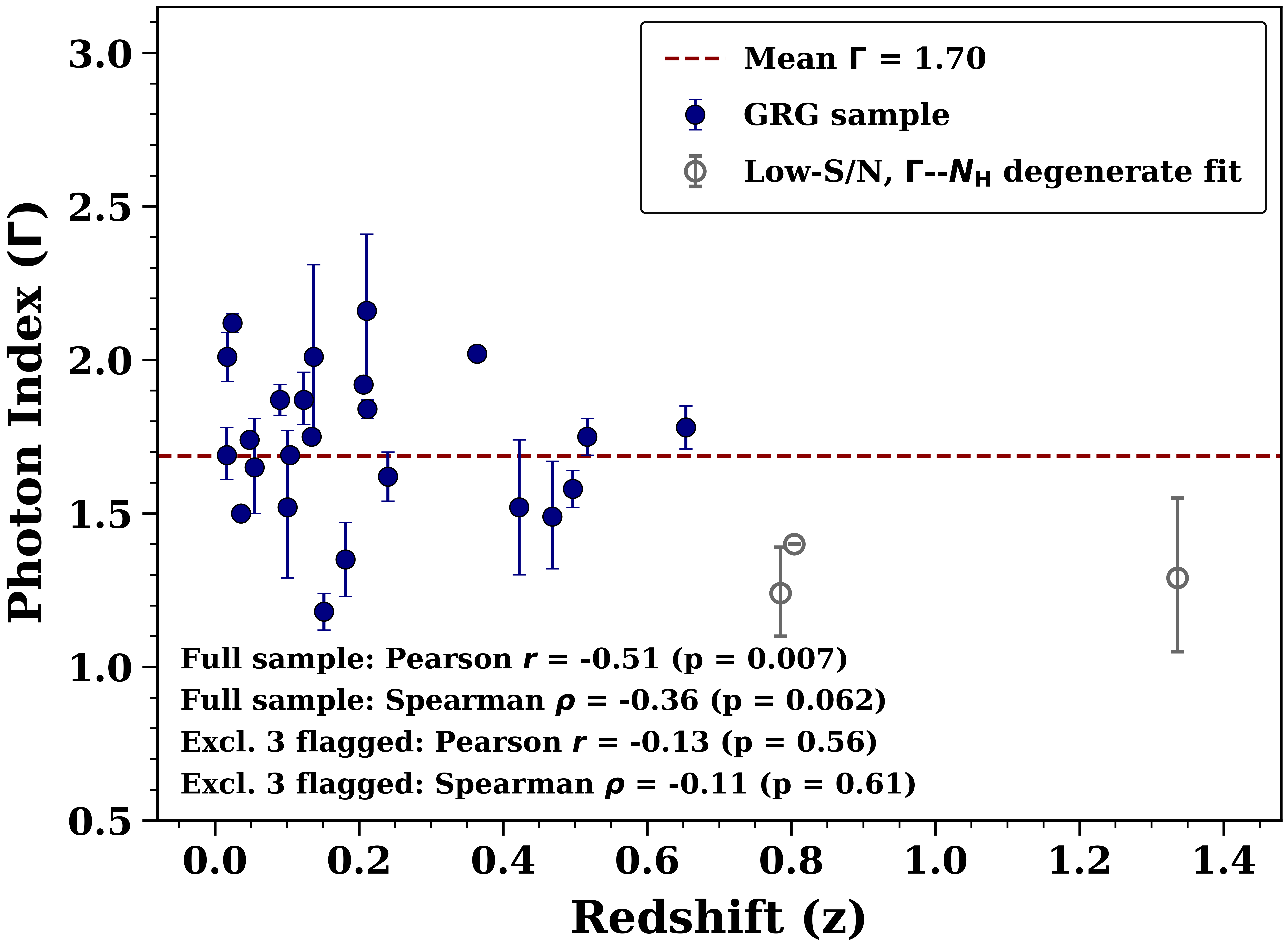}
\caption{Photon index $\Gamma$ as a function of redshift for the 27 GRGs in our sample. Open grey circles mark the three sources (Nos.~60, 61, 66) whose low $\Gamma$ values are statistically degenerate with higher intrinsic $N_{\rm H}$ at fixed $\Gamma\simeq1.7$ (Table~\ref{Table 10}). Once these three sources are excluded ($r=-0.13$, $p=0.56$; see main text). Correlation statistics are quoted for both the full sample and the sample excluding these three sources.}
\label{fig:PIvsRedshift}
\end{figure}

\begin{figure*}[t]
    \centering
    \includegraphics[width=1.0\linewidth]{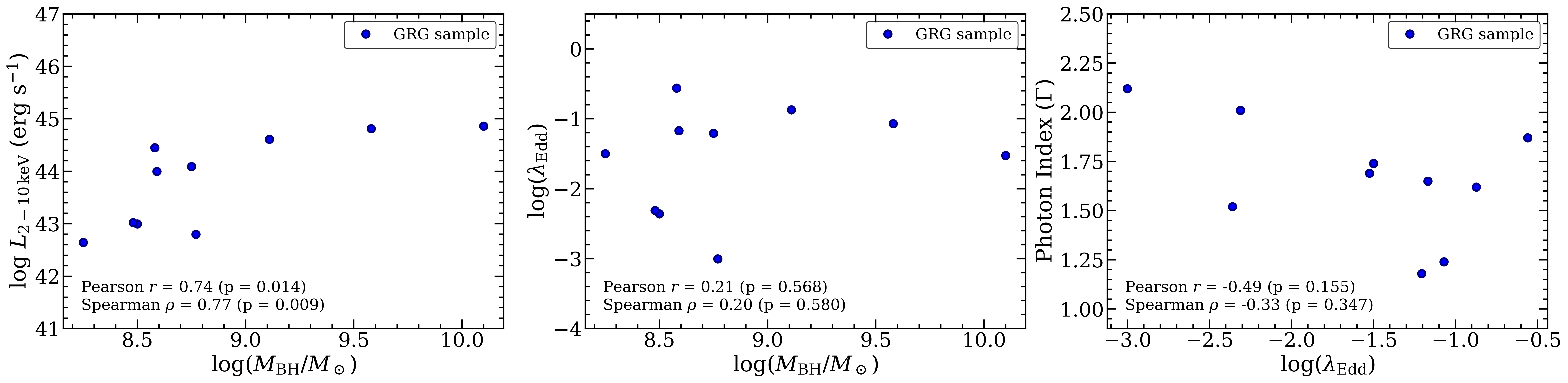}
    \caption{Correlations between black hole mass and X-ray properties of the GRG sample. Left: 2–10 keV X-ray luminosity vs. black hole mass, showing a positive correlation. Middle: Eddington ratio ($\lambda_{Edd}$) vs. black hole mass, with no clear dependence. Right: Photon index ($\Gamma$) vs. Eddington ratio, indicating a tentative negative trend. Red lines show linear fits, with Pearson and Spearman correlation coefficients reported in each panel.}
    \label{fig:Black Hole Mass Relations}
\end{figure*}

\subsection{Photon Index}

The photon index ($\Gamma$) of the X-ray continuum is a key diagnostic of AGN emission processes, probing the relative contributions of thermal Comptonisation in the accretion disk corona versus non-thermal emission from jets. To place the spectral properties of GRGs in context, we compared the $\Gamma$ distribution of our 27 GRGs with those of other AGN populations: radio-quiet Type 1 AGNs (46 sources) and Type 2 AGNs (31 sources) from \citet{2011A&A...530A..42C}, and radio-loud AGNs (26 sources) from \citet{1999ApJ...526...60S}. The resultant plot shown in Fig.~\ref{fig:enter-photon}. Our GRGs systematically exhibit flatter X-ray spectra than radio-quiet AGNs. Although the median $\Gamma$ for Type 2 AGNs is $\sim$1.9, closer to our GRGs values, it remains steeper by $\sim0.2$–0.3. Overall, GRGs appear to continue at least as hard as the larger radio-loud population, and harder than classical Type 1 and Type 2 AGNs.

The physical origin of these differences likely reflects both jet-related and accretion-linked components. In radio-quiet AGNs, typical slopes of $\Gamma\sim2.0$–2.1 are attributed to thermal Comptonisation in a hot corona above a radiatively efficient disk \citep{1991ApJ...380L..51H, 1993ApJ...413..507H, 2007A&ARv..15....1D}. Softer slopes trace cooler coronae at higher accretion rates, whereas harder spectra ($\Gamma\lesssim1.8$) often indicate non-thermal jet-linked emission. \citet{1999ApJ...526...60S} demonstrated that radio-loud AGNs in general have flatter spectra than radio-quiet ones, consistent with synchrotron or Synchrotron Self-Compton (SSC) contributions. Given that GRGs host large-scale radio jets, a dynamically important nuclear jet component is plausible and may contribute to their observed hard spectra.

\begin{table*}
    \centering
    \caption{List of GRGs with black hole mass and calculated Eddington luminosity ($L_{Edd}$) and Eddington ratio ($\lambda_{Edd}$).}
     \begin{tabular}{ccccccc}
\\ \hline
No. & IAU Name  & log ($M_{BH}/M_\odot$) & Ref. & $L_{\mathrm{Edd}}$ & $\lambda_{\mathrm{Edd}}$ &  \\
&    & & &($10^{47}$ erg) &  &  \\
\hline

19 & 2CXO J100601.7+345410 &  8.5 & \cite{refId0} & 0.39 & 0.0044 &  \\

60 & 2CXO J145307.9+221707 &  9.58 & \cite{2011ApJS..194...45S} & 4.79 & 0.0852   &  \\

37 & 2CXO J132834.1-012917 &  8.75 &  \cite{2004ApJ...614...91W}  & 0.71 & 0.0623    &  \\

%38 & 2CXO J010944.3+731157 & 0.181 & 8.58 &  \cite{2010AA...509A...6B}  &  &     &    \\

10 & 4XMM J233355.2-234340 &  8.25 & \cite{2022ApJS..261....5M} & 0.22 & 0.0317 &  \\

43 & 4XMM J174838.9-233520 &  9.11 &  \cite{Masetti} & 1.63 & 0.1341 &  \\

21 & 4XMM J204237.3+750802 &  10.1 & \cite{2001MNRAS.327.1111G} & 15.8 & 0.0299 & \\

5 & 4XMM J163232.1+823216 &  8.77 & \cite{2000ApJ...539L...9F} & 0.74 & 0.0010 & \\

29 & 4XMM J144851.0-400846 &  8.58 &  \cite{Masetti}  &  0.48  &  0.2757   &    \\

%2 & 4XMM J010724.9+322445 & 0.0167 & 9.59 & \cite{10.1093/mnras/staf055} & 4.90 & 0.0000001 & \\

%39 & 4XMM J172320.7+341758 & 0.206 & 8.01 & \cite{2006ApJ...637..669L} & & & \\

1 & 4XMM J005748.8+302109 & 9.32 & \cite{2021ApJ...908...19B} & 2.63 & 0.00002 & \\

32 & 4XMM J093952.7+355358 &  8.48 & \cite{2021MNRAS.500..215D} & 0.38 & 0.0049 &  \\

\hline

\end{tabular}
\label{Table 9}
\end{table*}

We also examined possible evolution of $\Gamma$ with redshift in the GRG sample (see Fig.~\ref{fig:PIvsRedshift}). Taking the sample at face value, a mild anti-correlation is found (Pearson $r=-0.51$, $p=0.007$; Spearman $\rho=-0.36$, $p=0.062$), which would suggest that higher-redshift GRGs tend to show flatter X-ray spectra, in contrast to the lack of $\Gamma$--$z$ evolution reported for Type~1 AGNs by \citet{2011A&A...530A..42C}.  However, closer inspection shows that this apparent trend is not robust, but is instead driven by a small number of low S/N spectra affected by the well known $\Gamma$--$N_{\rm H}$ degeneracy, as we demonstrate below.

Three sources dominate the high-redshift, low $\Gamma$ end of the trend: 4XMM~J133127.8+250049 ($z=0.804$), for which $\Gamma$ was fixed at 1.4 rather than fitted due to insufficient counts; 2CXO~J145307.9+221707 ($z=0.785$, dof$=38$), for which no intrinsic absorption was formally detected; and 4XMM~J235522.9+795518 ($z=1.336$, dof$=10$), the lowest count spectrum in our sample. Excluding these three sources, the Pearson and Spearman coefficients drop from $r=-0.51$ ($p=0.007$), $\rho=-0.36$ ($p=0.062$) for the full sample to $r=-0.13$ ($p=0.56$), $\rho=-0.11$ ($p=0.61$), i.e.\ the trend is no longer statistically significant.

Given this degeneracy, we tested it directly by re-fitting all three spectra with $\Gamma$ frozen at 1.7 (close to the sample median) and $N_{\rm H}$ left free (Table~\ref{Table 10}). In every case an equally good fit is obtained, with the intrinsic column density increasing correspondingly -- most strikingly for 2CXO~J145307.9+221707, where a previously undetected absorber becomes significant ($N_{\rm H}=1.37\pm0.61\times10^{22}$~cm$^{-2}$) once $\Gamma$ is fixed near the sample median, and for 4XMM~J235522.9+795518, where an F-test formally confirms that the free $\Gamma$ and fixed $\Gamma$ solutions are statistically indistinguishable ($F=2.48$, $p=0.15$). We therefore conclude that the apparent flattening of $\Gamma$ with redshift is driven by unresolved $\Gamma$--$N_{\rm H}$ degeneracy in a handful of low S/N spectra rather than by genuine spectral evolution, and we regard the $\Gamma$--$z$ trend reported in this work as tentative, pending confirmation with higher S/N data for a larger, more uniform sample.

From the unification perspective, the similarity of GRG spectral slopes to those of other radio-loud AGNs is instructive. Standard orientation-based schemes \citep{1995PASP..107..803U} predict that Type 1/2 differences primarily arise from obscuration, not intrinsic slope changes. The fact that GRGs cluster at the hard end of the radio-loud population suggests they are not a distinct “soft-spectrum” class, but rather consistent with being large-scale, evolved radio galaxies whose nuclear X-ray spectra remain dominated by jet-linked or inefficient accretion processes.

\subsection{Preliminary Trends in Black Hole Mass Relationships:}

For the 10 GRGs in our sample (see Table~\ref{Table 9}) with available black hole mass estimates, we examined possible links between $M_{\rm BH}$, X-ray luminosity, and accretion efficiency (Fig.~\ref{fig:Black Hole Mass Relations}). The Eddington luminosity \citep{1926ics..book.....E, 1995A&A...303L...1L} was computed as

\begin{equation}
L_{\rm Edd} = 1.26 \times 10^{38} \left( \frac{M_{\rm BH}}{M_\odot} \right) \, {\rm erg \, s^{-1}} ,
\end{equation}

and the Eddington ratio \citep[e.g.,][]{2006ApJ...648..128K} was then defined by  

\begin{equation}
\lambda_{\rm Edd} = \frac{L_{\rm bol}}{L_{\rm Edd}}.
\end{equation}

These tests are necessarily tentative given the small number of sources and the heterogeneous origin of the mass estimates, but they provide a first look at how central engine properties may connect to X-ray emission in GRGs.

We find a positive correlation between $\log M_{\rm BH}$ and $\log L_{2-10\,{\rm keV}}$ (Pearson $r=0.74$, $p=0.014$), which implies that the most massive black holes in the sample also tend to be the most X-ray luminous. This is consistent with the expectation that nuclear power, whether from accretion or jet-linked processes, broadly scales with black hole mass.

By contrast, there is no significant correlation between $\log M_{\rm BH}$ and $\log(L_{\rm bol}/L_{\rm Edd})$. This indicates that accretion efficiency is not directly set by black hole mass in GRGs, but more likely determined by fueling conditions or environment. This agrees with previous findings that GRGs typically accrete at very low fractions of the Eddington rate \citep{2020A&A...642A.153D}.

Finally, we see a weak negative trend between the photon index $\Gamma$ and $\log(L_{\rm bol}/L_{\rm Edd})$ (Pearson $r=-0.49$, $p=0.16$). Although not statistically significant, this behavior is qualitatively in line with the suggestion that GRGs host radiatively inefficient accretion flows (RIAFs) \citep{2003ApJ...592.1042I}, where higher accretion rates may still correspond to relatively hard X-ray spectra. This contrasts with the positive $\Gamma$–$L/L_{\rm Edd}$ relation observed in luminous quasars \citep{2006ApJ...646L..29S, 2008ApJ...682...81S}.

We emphasize, however, that these results are based on only 10 sources. With such small statistics, the derived correlations should be regarded as illustrative rather than definitive, and the apparent trends could be affected by outliers, intrinsic scatter, or systematic uncertainties in the mass estimates. A larger, homogeneous sample with reliable black hole masses and deep X-ray coverage will be required to confirm whether GRGs consistently follow these trends or whether their behaviour departs from that of other radio-loud AGN.

In above discussed trends, we reiterate a caveat on the sample selection : the requirement of S/N~$>$~20 for reliable spectral fitting preferentially retains X-ray brighter, and hence typically nearer and/or more core luminous, GRGs. This is likely to bias our results in specific ways. First, the detected fraction of intrinsically absorbed sources and the incidence and equivalent widths of Fe K$\alpha$ emission reported above may be overestimated for some sources relative to the true GRG population, since faint, heavily obscured, or weak core GRGs are less likely to satisfy the S/N threshold and are thus preferentially excluded. Second, the median photon index, the presented co-relations, and the black hole mass trends discussed above should be regarded as characteristic of this X-ray selected, relatively nearby and radio-core luminous subset of GRGs, rather than of the GRG population as a whole. Deeper, more sensitive X-ray observations of currently excluded GRGs will be needed to test whether these trends persist across the full GRG population.

\section{Conclusions}
\label{conclusion}

The main conclusions from the comprehensive study of GRGs using the \textit{XMM-Newton}, \textit{Chandra} X-ray and radio  data are as follows:
\begin{enumerate}
    \item This is the first systematic X-ray spectral study of 27 GRGs using archival data from \textit{XMM-Newton} and \textit{Chandra} observatories.\\
    
    \item We found $\sim$44\% of the GRG sample shows significant intrinsic X-ray absorption ($N_{\mathrm{H,int}} \gtrsim 10^{21}$~cm$^{-2}$), suggesting the presence of obscuring circumnuclear material.\\
    
    \item Narrow Fe K$\alpha$ emission lines were detected in several GRGs, with equivalent widths up to $\sim$570~eV, consistent with cold, optically thick reflection.\\
    
    \item A variety of soft X-ray components were observed, including thermal plasma emission (kT $\sim 0.2$--$0.8$~keV) and features consistent with warm absorbers and ionized outflows.\\
    
    \item In several cases, models incorporating ionized smeared absorption indicated the presence of high-velocity winds near the active nucleus.\\

    %\item Our analysis confirms the diversity in the nuclear environments of GRGs, and supports scenarios involving clumpy tori, AGN feedback mechanisms, and episodic or restarting nuclear activity.\\

    %\item The findings highlight the critical role of multi-phase X-ray diagnostics in understanding the accretion and feedback processes in giant radio galaxies.\\
    
    \item The photon indices ($\Gamma$) of the GRG nuclei with a median of $\sim$1.6, are in line with those of other radio-loud AGNs, but with a tendency for flatter spectra at higher redshift.\\
    
    \item Many GRGs exhibit stronger nuclear X-ray luminosity relative to their extended radio lobe power, which may indicate either restarted AGN activity.\\
    
    \item For the 10 sources with published black hole mass estimates, we examined preliminary relations between $M_{\rm BH}$, X-ray luminosity, Eddington ratio, and photon index. A positive trend between $M_{\rm BH}$ and $L_{2-10\,{\rm keV}}$ is seen, while no significant dependence of Eddington ratio on $M_{\rm BH}$ is found. A weak negative relation between $\lambda_{\rm Edd}$ and $\Gamma$ is consistent with radiatively inefficient accretion scenarios, but the limited sample prevents firm conclusions.

\end{enumerate}

\section*{CRediT authorship contribution statement}

\textbf{Niraj Maurya} : Conceptualization, Data Curation, Formal analysis, Investigation, Writing – original draft, Methodology, Software; \textbf{Satish S. Sonkamble} : Conceptualization, Data curation, Formal analysis, Investigation, Methodology, Supervision, Validation, Writing – original draft, Writing – review \& editing; \textbf{Suraj Dhiwar} : data curation, Formal analysis; \textbf{Mahadev B. Pandge} : Writing – review \& editing, Funding acquisition, Project administration, Supervision; \textbf{S. Ilani Loubser} : Writing – review \& editing; \textbf{Avinash Kale} : Writing – review \& editing.

\section*{Data Availability}
We have made use of archival data from observations with the European Photon Imaging Camera (EPIC) onboard the {\it XMM$-$Newton} observatory and the Giant Meterwave Radio Telescope (GMRT). All X-ray data were obtained from the HEASARC archive and are publicly accessible via \url{https://heasarc.gsfc.nasa.gov/cgi-bin/W3Browse/w3browse.pl}

\section*{Declaration of competing interest}
We declare no conflict of interest.

\section*{Acknowledgements}
NM and MBP gratefully acknowledge the support from the Science and Engineering Research Board (SERB), New Delhi,  under the `SERB CRG’ funding with sanction no. CRG/2023/003463. MBP gratefully acknowledges the support from IUCAA under the Visiting Associate for college and university faculties. SIL is supported in part by the National Research Foundation (NRF) of South Africa (CPRR240414214079). Any opinion, finding, and conclusion or recommendation expressed in this material is that of the author(s), and the NRF does not accept any liability in this regard. This research has made use of the data from the HEASARC Archive. This research has made use of  NASA's  Astrophysics Data  System and of the  NASA/IPAC  Extragalactic Database (NED), which is operated by the Jet  Propulsion Laboratory, California Institute of Technology, under contract with the National Aeronautics and Space  Administration.
Facilities: HEASARC, \texttt{XSPEC}.

\def\nat{Nature}%
\def\aj{AJ}%
\def\actaa{Acta Astron.}% % Acta Astronomica
\def\araa{ARA\&A}% % Annual Review of Astron and Astrophys
\def\apj{ApJ}% % Astrophysical Journal
\def\apjl{ApJ}% % Astrophysical Journal, Letters
\def\apjs{ApJS}% % Astrophysical Journal, Supplement
\def\aap{A\&A}% % Astronomy and Astrophysiof
\def\aapr{A\&A~Rev.}% % Astronomy and Astrophysics Reviews
\def\aaps{A\&AS}% % Astronomy and Astrophysics, Supplement
\def\apss{Ap\&SS}% %Astrophysics and Space Science
\def\baas{BAAS}% % Bulletin of the AAS
\def\caa{Chinese Astron. Astrophys.}% % Chinese Astronomy and Astrophysics
\def\cjaa{Chinese J. Astron. Astrophys.}% % Chinese Journal of Astronomy and Astrophysics
\def\icarus{Icarus}% % Icarus
\def\jcap{J. Cosmology Astropart. Phys.}% % Journal of Cosmology and Astroparticle Physics
\def\jrasc{JRASC}% % Journal of the RAS of Canada
\def\memras{MmRAS}% % Memoirs of the RAS
\def\mnras{MNRAS}% % Monthly Notices of the RAS
\def\na{New A}% % New Astronomy
\def\nar{New A Rev.}% % New Astronomy Review
\def\pra{Phys.~Rev.~A}% % Physical Review A: General Physics
\def\prb{Phys.~Rev.~B}% % Physical Review B: Solid State
\def\prc{Phys.~Rev.~C}% % Physical Review C
\def\prd{Phys.~Rev.~D}% % Physical Review D
\def\pre{Phys.~Rev.~E}% % Physical Review E
\def\prl{Phys.~Rev.~Lett.}% % Physical Review Letters
\def\pasa{PASA}% % Publications of the Astron. Soc. of Australia
\def\pasp{PASP}% % Publications of the ASP
\def\pasj{PASJ}% 
\def\sovast{SOVAST}% 
\def\ssr{Space Sci. Rev.}
\def\physrep{Phys. Rep.}



\bibliographystyle{elsarticle-harv} 
\bibliography{grg}

\appendix
\section{Best-fit X-ray Spectrum of GRGs}
\label{app}
\begin{figure*}[p]
\centering

%---------------- Row 1 ----------------%
\begin{subfigure}[t]{0.52\textwidth}
    \centering
    \includegraphics[width=\linewidth]{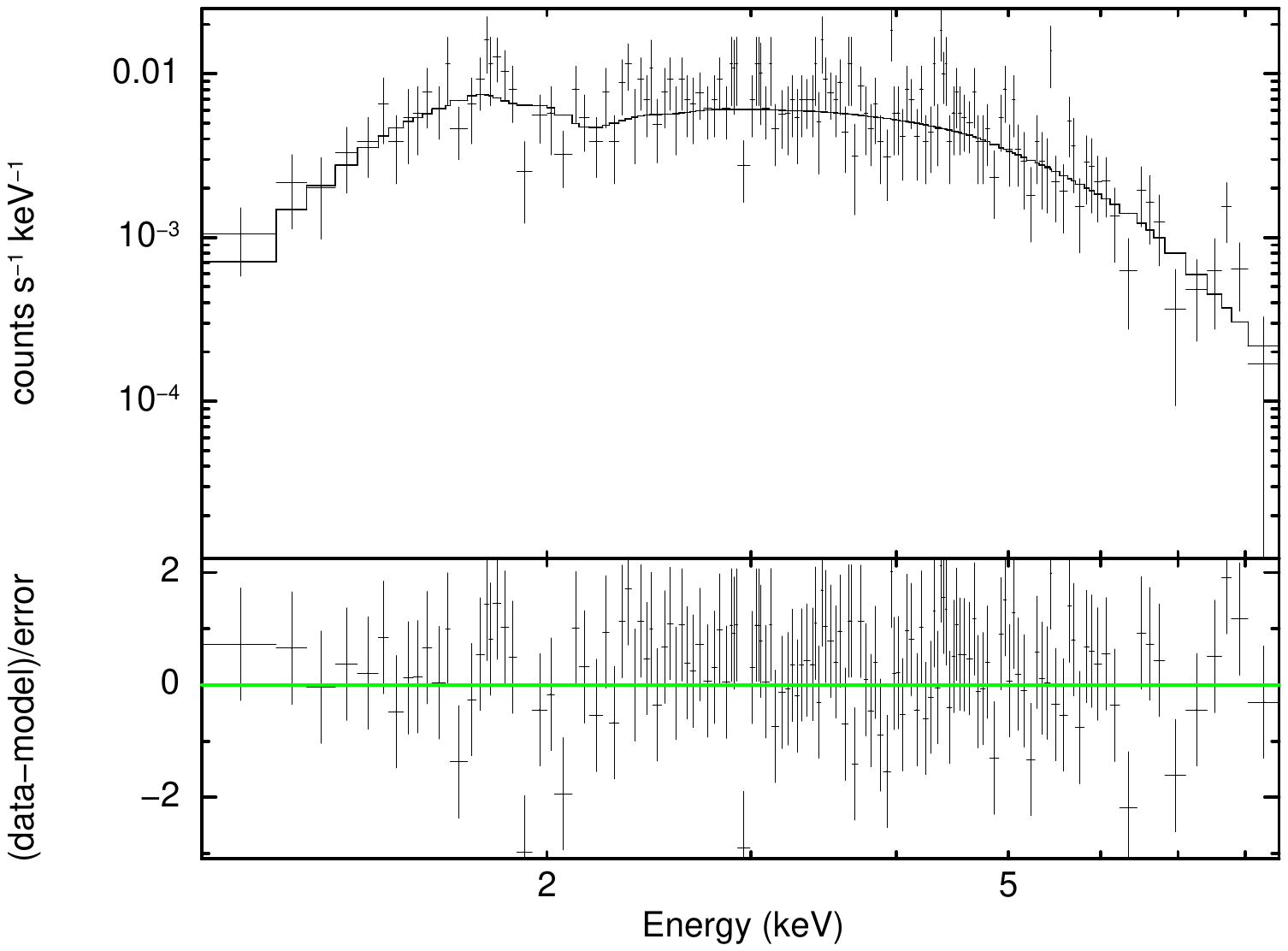}
    \caption{Model 1: 2CXO J042925.8+003304}
    \label{fig:a}
\end{subfigure}
\hspace{-1cm}
\begin{subfigure}[t]{0.52\textwidth}
    \centering
    \includegraphics[width=\linewidth]{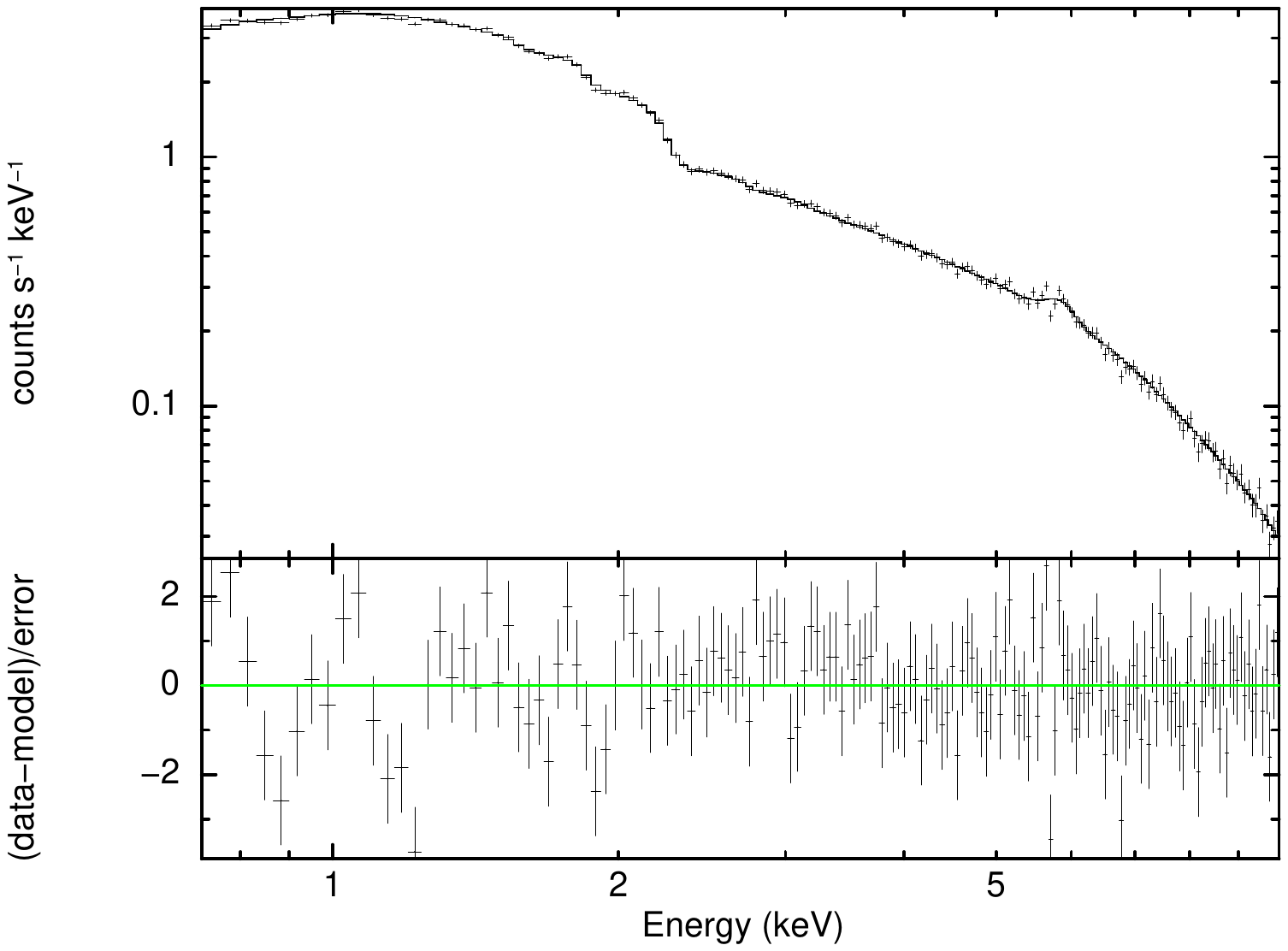}
    \caption{Model 2: 4XMM J204237.3+750802}
    \label{fig:b}
\end{subfigure}

\vspace{0.5cm}

%---------------- Row 2 ----------------%
\begin{subfigure}[t]{0.52\textwidth}
    \centering
    \includegraphics[width=\linewidth]{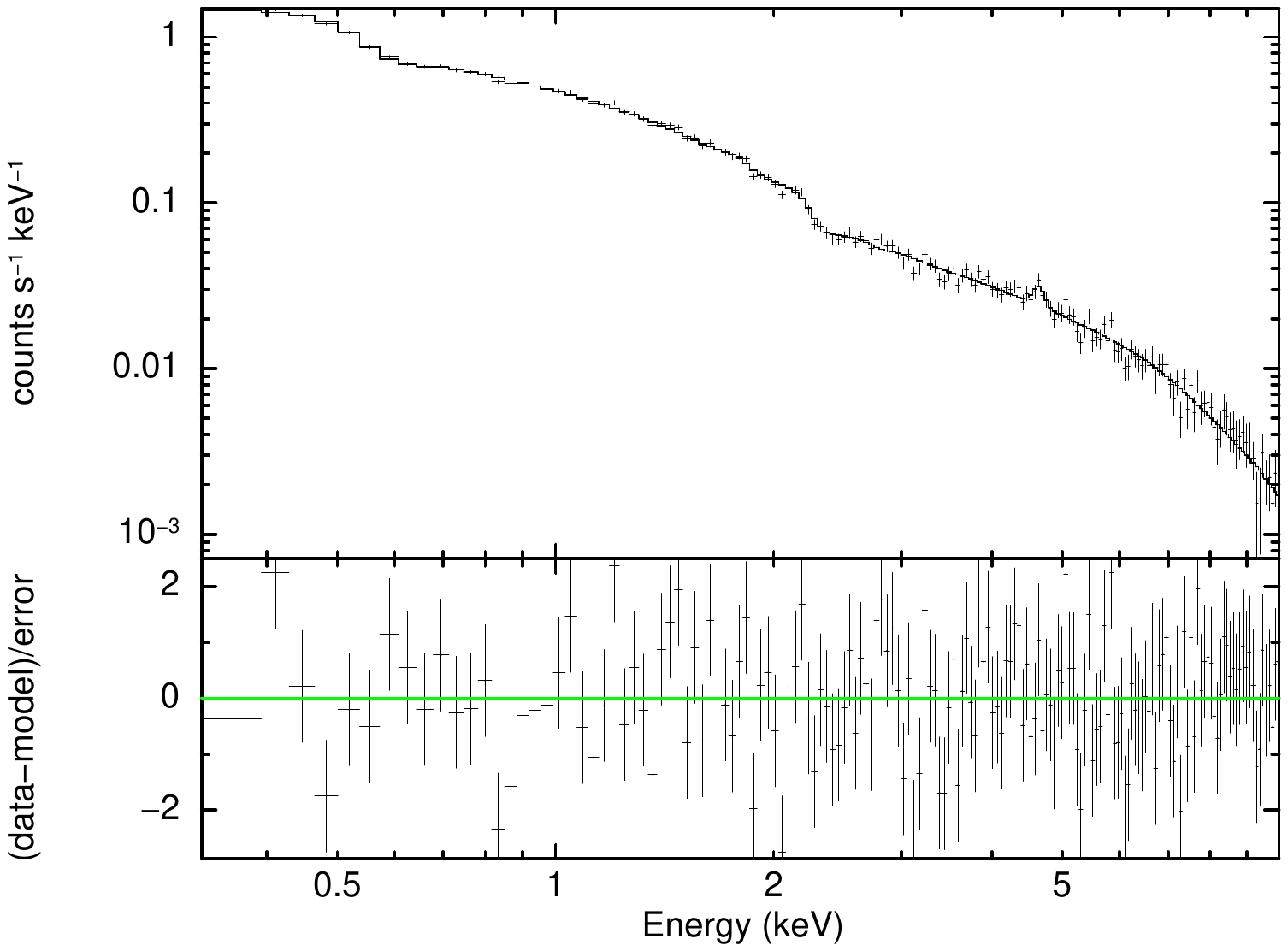}
    \caption{Model 3: 4XMM J142735.6+263214}
    \label{fig:c}
\end{subfigure}
\hspace{-1cm}
\begin{subfigure}[t]{0.52\textwidth}
    \centering
    \includegraphics[width=\linewidth]{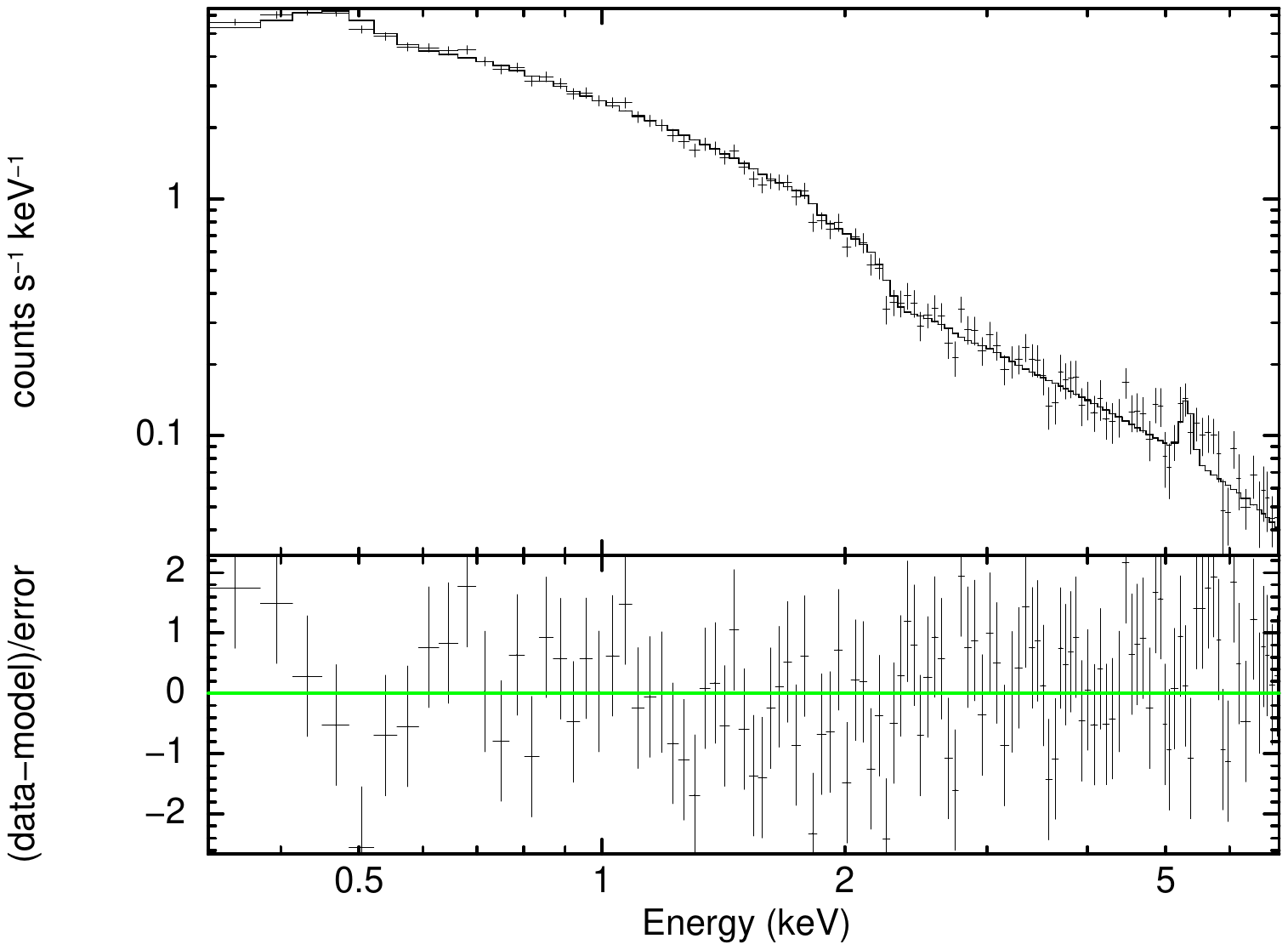}
    \caption{Model 4: 4XMM J172320.7+341758}
    \label{fig:d}
\end{subfigure}

\vspace{0.5cm}

%---------------- Row 3 ----------------%
\begin{subfigure}[t]{0.52\textwidth}
    \centering
    \includegraphics[width=\linewidth]{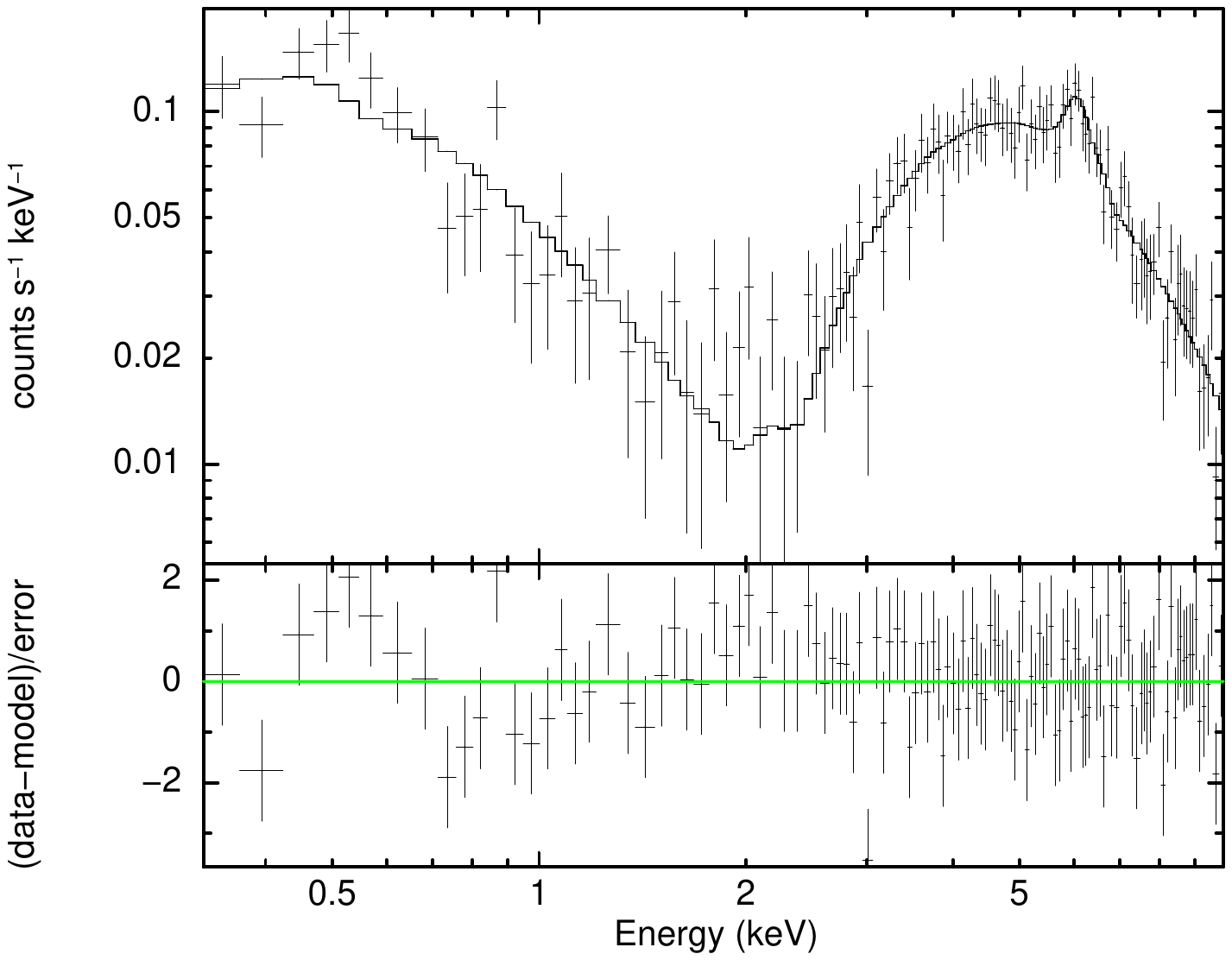}
    \caption{Model 5: 4XMM J162804.0+514631}
    \label{fig:e}
\end{subfigure}
\hspace{-1cm}
\begin{subfigure}[t]{0.52\textwidth}
    \centering
    \includegraphics[width=\linewidth]{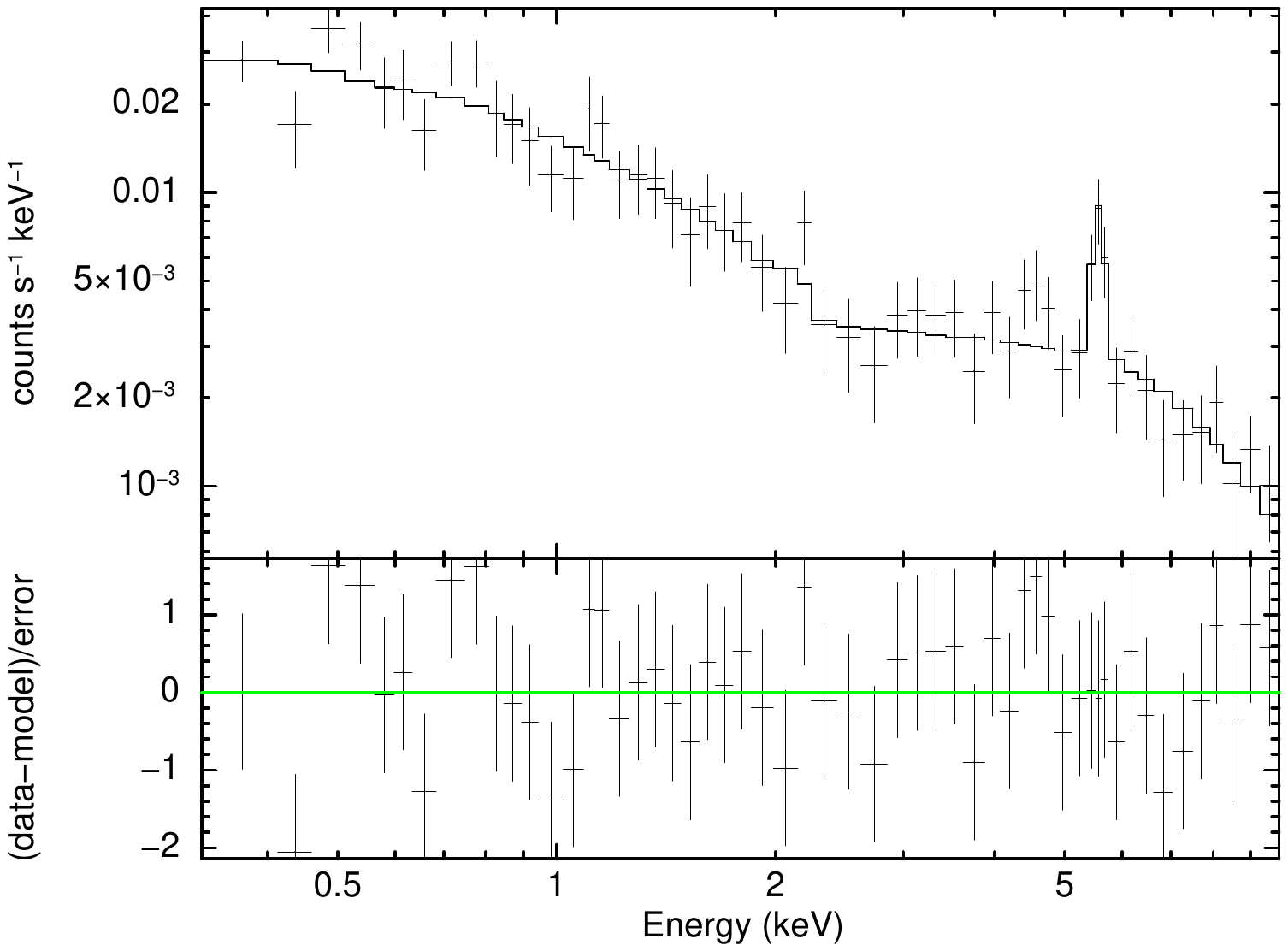}
    \caption{Model 6: 4XMM J093952.7+355358}
    \label{fig:f}
\end{subfigure}

\caption{Representative X-ray spectra of some GRGs from our sample and their residuals are shown in bottom panels.}
\label{fig:sixpanel}

\end{figure*}

Figure~\ref{fig:sixpanel} shows the best-fit X-ray spectra of representative GRGs from each of the six spectral models (Model 1 to Model 6) discussed in Section~\ref{Analysis}. In each panel, the upper part shows the data points along with the best-fitting model (solid line), while the lower part shows the residuals from the best-fitting model in units of $\sigma$. The best-fitting parameters of the corresponding models are listed in Tables~\ref{Table 1}--\ref{Table 6}.

\captionsetup[subfigure]{font=normalsize}

\section{Table}
Supporting tables related to the X-ray spectral analysis.

\begin{table*}
    \centering
    \caption{Fe K$\alpha$ equivalent widths (detections and 90\%
    upper limits) and 2--10~keV luminosities for the full sample of
    27 GRGs. Upper limits (marked $<$) are to be derived by adding a
    narrow ($\sigma=10$~eV) Gaussian fixed at rest-frame 6.4~keV to
    the best-fit continuum and stepping its normalization to
    $\Delta\chi^{2}=2.7$.}
    \begin{tabular}{clccc}
    \hline
    No. & IAU Name & Fe K$\alpha$ EW (eV) & $\log L_{2-10\,\rm keV}$ (erg s$^{-1}$) & Detection (Y/N) \\
    \hline
    21  & 4XMM J204237.3+750802 & $118.53^{+23.01}_{-21.63}$   & 44.86 & Y \\
    43  & 4XMM J174838.9-233520 & $200.00^{+67.53}_{-54.42}$   & 44.61 & Y \\
    18  & 4XMM J031819.1+682932 & $<91.48$ & 44.31 & N \\
    2  &  4XMM J010724.9+322445 & -- & 40.85 & -- \\
    46  & 4XMM J142735.6+263214 & $63.98^{+21.11}_{-15.39}$    & 44.94 & Y \\
    11 & 4XMM J162804.0+514631 & $218.48^{+75.88}_{-71.15}$   & 44.00 & Y \\
    5 & 4XMM J163232.1+823216 &  $61.29^{+31.07}_{-40.04}$ & 42.80 & Y \\
    48 & 4XMM J123526.6+212034 & -- & 42.11 & -- \\
    61 & 4XMM J133127.8+250049 & $391.56^{+387.03}_{-318.90}$ & 44.38 & Y \\
    1 & 4XMM J005748.8+302109 & -- & 41.60 & -- \\
    39 & 4XMM J172320.7+341758 & $151.01^{+68.89}_{-61.40}$ & 44.96 & Y \\
    10 & 4XMM J233355.2-234340 & $<46.28$ & 42.64 & N \\
    58 & 4XMM J121952.3+472058 & -- & 43.42 & -- \\
    41 & 4XMM J225336.0-345530 & $<321.21$ & 43.21 & N \\
    31 & 4XMM J031301.9+412001 & $151.69^{+83.53}_{-70.98}$ & 44.15 & Y \\
    8 & 4XMM J131217.0+445021 & -- & 41.00 & -- \\
    40 & 4XMM J013929.8+395712 & -- & 42.04 & -- \\
    55 & 4XMM J152311.0+520303 & -- & 44.12 & -- \\
    66 & 4XMM J235522.9+795518 & -- & 44.96 & -- \\
    32 & 4XMM J093952.7+355358 & $567.66^{+179.67}_{-223.16}$ & 43.02 & Y \\
    29 & 4XMM J144851.0-400846 & $85.69^{+25.86}_{-15.90}$    & 44.45 & Y \\
    38 & 2CXO J010944.3+731157 & $<211.22$ & 45.67 & N \\
    19 & 2CXO J100601.7+345410 & $224.63^{+257.07}_{-218.67}$ & 43.00 & Y \\
    60 & 2CXO J145307.9+221707 & $<346.64$ & 44.81 & N \\
    54 & 2CXO J114720.7-125309 & $<218.47$ & 44.91 & N \\
    37 & 2CXO J132834.1-012917 & $<149.42$ & 44.09 & N \\
    53 & 2CXO J042925.8+003304 & $<222.98$ & 44.72 & N \\
    \hline
    \end{tabular}
    \label{Table FeK}
\end{table*}

\begin{table*}
    \centering
    \caption{$F$-test results for the statistical significance
    of the absorption edges (\texttt{zedge}) and smeared ionized
    absorber (\texttt{swind1}) components relative to the baseline
    model (\texttt{phabs*zwabs*zpo}). $\Delta\chi^{2}$ and
    $\Delta$dof are computed between the baseline and adopted models
    for the same source and dataset.}
    \begin{tabular}{clcccccc}
    \hline
    No. & Source & Model & $\chi^{2}_{\rm base}/{\rm dof}_{\rm base}$ & $\chi^{2}_{\rm model}/{\rm dof}_{\rm model}$ & $\Delta\chi^{2}$ (dof) & $F$ & $p_F$ \\
    \hline
    18 & 4XMM J031819.1+682932 & Model 2 (edge)         & 131/111 & 113/109 &  18 (2)  &  8.7  & $3.2\times10^{-4}$ \\
    43 & 4XMM J174838.9-233520 & Model 2 (edge+line)    & 157/117 & 109/110 &  48 (7)  &  6.9  & $8.0\times10^{-7}$ \\
    21 & 4XMM J204237.3+750802 & Model 2 (edge+line)    & 331/165 & 197/158 &  134 (7) &  15.3 & $3.0\times10^{-15}$ \\
    5  & 4XMM J163232.1+823216 & Model 3 (swind1+line) & 181/140 & 132/136 &  49 (4)  &  12.6 & $9.2\times10^{-9}$ \\
    29 & 4XMM J144851.0-400846 & Model 3 (swind1+line)  & 310/155 & 154/151 & 156 (4) &  38.2 & $<10^{-15}$ \\
    46 & 4XMM J142735.6+263214 & Model 3 (swind1+line)  & 369/160 & 167/156 & 202 (4) &  47.2 & $<10^{-15}$ \\
    \hline
    \end{tabular}
    \label{Table Ftest}
\end{table*}

\begin{table*}
\small
\caption{Refits of the three lowest S/N, low $\Gamma$ sources with the photon index frozen at $\Gamma=1.7$ (i.e. sample median) and intrinsic $N_{\rm H}$ left free, compared to the original free parameter fits reported in Table~\ref{Table 1}. $^{f}$ - fixed the parameter.}
\centering
\begin{tabular}{lcccccc}
\hline
No. & Source & $z$ & \multicolumn{2}{c}{Original fit} & \multicolumn{2}{c}{Refit ($\Gamma=1.7^{f}$)} \\
 & & & $\Gamma$, $N_{\rm H}$ (10$^{22}$ cm$^{-2}$) & $\chi^2$/dof (prob.) & $N_{\rm H}$ (10$^{22}$ cm$^{-2}$) & $\chi^2$/dof (prob.) \\
\hline
66 & 4XMM J235522.9+795518   & 1.336 & $1.29$, $10.72$ & $8/10$ (0.63) & $16.30\pm2.43$ & $9.98/11$ (0.53) \\
61 & 4XMM J133127.8+250049   & 0.804 & $1.4^{f}$, $0.13$ & $30/28$ (0.36) & $0.26\pm0.08$ & $35.69/28$ (0.15) \\
60 & 2CXO J145307.9+221707   & 0.785 & $1.24$, $\approx0$ & $38/38$ (0.47) & $1.37\pm0.61$ & $41.98/38$ (0.30) \\
\hline
\end{tabular}
\label{Table 10}
\end{table*}



%% For example:

%% \bibitem[Aladro et al.(2015)]{Aladro15} Aladro, R., Martín, S., Riquelme, D., et al. 2015, \aas, 579, A101

%%\end{thebibliography}

\end{document}